\PassOptionsToPackage{usenames,dvipsnames}{xcolor}
\documentclass[twocolumn,twocolappendix]{aastex631}
\pdfoutput=1 
\usepackage[T1]{fontenc}
\usepackage{ae,aecompl}
\usepackage[utf8]{inputenc}
\usepackage[english]{babel}
\usepackage{amsmath,amstext}
\usepackage{apjfonts}
\usepackage{wasysym}
\usepackage{chngcntr}
\usepackage[figure,figure*,table,table*]{hypcap}
\usepackage[hang]{footmisc}
\usepackage{wasysym}
\graphicspath{{./}{figures/}}
\input{hyperlink-year-only-natbib-patch}
\allowdisplaybreaks
\graphicspath{{figures/}}

\newcommand*{\https}[1]{\href{https://#1}{\nolinkurl{#1}}}
\newcommand*{\http}[1]{\href{http://#1}{\nolinkurl{#1}}}

\defcitealias{Geha2017}{{Paper\,I}}
\defcitealias{Mao2021}{{Paper\,II}}
\defcitealias{saga3:Mao2024}{{Paper\,III}}
\defcitealias{saga4:Geha2024}{{Paper\,IV}}
\defcitealias{saga5:Wang2024}{{Paper\,V}}

\newcommand*{\paperthree}{\citetalias{saga3:Mao2024}}
\newcommand*{\paperfour}{\citetalias{saga4:Geha2024}}

\newcommand*{\msun}{\ensuremath{\text{M}_\odot}}
\newcommand*{\mstar}{\ensuremath{M_{\star}}}

\shorttitle{The SAGA Survey.\ V.}
\shortauthors{Wang et al.}

\begin{document}
\title{The SAGA Survey.\ V.\ Modeling Satellite Systems around Milky Way-mass Galaxies with Updated UniverseMachine}

\author[0000-0001-8913-626X]{Yunchong Wang}
\affiliation{Kavli Institute for Particle Astrophysics and Cosmology and Department of Physics, Stanford University, Stanford, CA 94305, USA}
\affiliation{SLAC National Accelerator Laboratory, Menlo Park, CA 94025, USA}
\author[0000-0002-1182-3825]{Ethan~O.~Nadler}
\affiliation{Carnegie Observatories, 813 Santa Barbara Street, Pasadena, CA 91101, USA}
\affiliation{Department of Physics $\&$ Astronomy, University of Southern California, Los Angeles, CA, 90007, USA}
\affiliation{Department of Astronomy \& Astrophysics, University of California, San Diego, La Jolla, CA 92093, USA}
\author[0000-0002-1200-0820]{Yao-Yuan~Mao}
\affiliation{Department of Physics and Astronomy, University of Utah, Salt Lake City, UT 84112, USA}
\author[0000-0003-2229-011X]{Risa~H.~Wechsler}
\affiliation{Kavli Institute for Particle Astrophysics and Cosmology and Department of Physics, Stanford University, Stanford, CA 94305, USA}
\affiliation{SLAC National Accelerator Laboratory, Menlo Park, CA 94025, USA}
\author[0000-0002-5969-1251]{Tom Abel}
\affiliation{Kavli Institute for Particle Astrophysics and Cosmology and Department of Physics, Stanford University, Stanford, CA 94305, USA}
\affiliation{SLAC National Accelerator Laboratory, Menlo Park, CA 94025, USA}
\author[0000-0002-2517-6446]{Peter Behroozi}
\affiliation{Steward Observatory and Department of Astronomy, University of Arizona, Tuscon, AZ 85721, USA}
\affiliation{Division of Science, National Astronomical Observatory of Japan, 2-21-1 Osawa, Mitaka, Tokyo 181-8588, Japan}
\author[0000-0002-7007-9725]{Marla~Geha}
\affiliation{Department of Astronomy, Yale University, New Haven, CT 06520, USA}
\author[0000-0002-8320-2198]{Yasmeen~Asali}
\affiliation{Department of Astronomy, Yale University, New Haven, CT 06520, USA}
\author[0000-0002-4739-046X]{Mithi~A.~C.~de~los~Reyes}
\affiliation{Department of Physics and Astronomy, Amherst College, Amherst, MA 01002}
\author[0000-0002-0332-177X]{Erin~Kado-Fong}
\affiliation{Department of Physics and Yale Center for Astronomy \& Astrophysics, Yale University, New Haven, CT 06520, USA}
\author[0000-0002-3204-1742]{Nitya~Kallivayalil}
\affiliation{Department of Astronomy, University of Virginia, Charlottesville, VA 22904, USA}
\author[0000-0002-9599-310X]{Erik~J.~Tollerud}
\affiliation{Space Telescope Science Institute, Baltimore, MD 21218, USA}
\author[0000-0001-6065-7483]{Benjamin~Weiner}
\affiliation{Steward Observatory and Department of Astronomy, University of Arizona, Tuscon, AZ 85721, USA}
\author[0000-0002-5077-881X]{John~F.~Wu}
\affiliation{Space Telescope Science Institute, Baltimore, MD 21218, USA}
\affiliation{Center for Astrophysical Sciences, Johns Hopkins University, Baltimore, MD 21218, USA}

\correspondingauthor{Yunchong Wang}
\email{ycwang19@stanford.edu}

\begin{abstract}
Environment plays a critical role in shaping the assembly of low-mass galaxies. Here, we use the \textsc{UniverseMachine} (UM) galaxy--halo connection framework and the Data Release 3 of the Satellites Around Galactic Analogs (SAGA) Survey to place dwarf galaxy star formation and quenching into a cosmological context. UM is a data-driven forward model that flexibly parameterizes galaxy star formation rates (SFR) using only halo mass and assembly history. We add a new quenching model to UM, tailored for galaxies with $\mstar\lesssim 10^9\msun$, and constrain the model down to $\mstar \gtrsim 10^7 \msun$ using new SAGA observations of 101 satellite systems around Milky Way (MW)-mass hosts and a sample of isolated field galaxies in a similar mass range from the Sloan Digital Sky Survey (SDSS). The new best-fit model, `UM-SAGA,' reproduces the satellite stellar mass functions, average SFRs, and quenched fractions in SAGA satellites while keeping isolated dwarfs mostly star forming. The enhanced quenching in satellites relative to isolated field galaxies leads the model to maximally rely on halo assembly to explain the observed environmental quenching. Extrapolating the model down to $\mstar\sim 10^{6.5}\msun$ yields a quenched fraction of $\gtrsim 30\%$ for isolated field galaxies and $\gtrsim 80\%$ for satellites of MW-mass hosts at this stellar mass. Spectroscopic surveys can soon test this specific prediction to reveal the relative importance of internal feedback, cessation of mass and gas accretion, satellite-specific gas processes, and reionization for the evolution of faint low-mass galaxies.
\end{abstract}

\keywords{
 \href{http://astrothesaurus.org/uat/416}{Dwarf galaxies (416)}, \href{http://astrothesaurus.org/uat/595}{Galaxy formation (595)}, \href{http://astrothesaurus.org/uat/2040}{Galaxy quenching (2040)},
 \href{http://astrothesaurus.org/uat/1083}{N-body simulations (1083)},
 \href{http://astrothesaurus.org/uat/1880}{Galaxy dark matter halos (1880)}
}

\section{Introduction}
\label{sec:intro}

Low-mass galaxies ($\mstar \lesssim 10^9\msun$) have become a crucial test of both hierarchical structure formation and dark matter models due to their old stellar populations, their low baryonic content, and their sensitivity to the impact of changes to the small-scale power spectrum (see, e.g., \citealt{2017ARA&A..55..343B,2022NatAs...6..897S} for reviews). Due to their intrinsic faintness, most of the known low-mass galaxies, especially the ultra-faint-dwarfs (UFDs, $\mstar<10^5\msun$), which are the faintest observable galaxies, have mainly been observed within or in the vicinity of the dark matter halos of the Milky Way (MW) and M31, a.k.a. the Local Group \citep[LG; for a review, see][]{2019ARA&A..57..375S}. 

Low-mass galaxies demonstrate diverse star formation properties in different cosmic environments. The low-mass satellite galaxies of the MW and M31 are mostly quenched of their star formation~\citep{2012AJ....144....4M,2013AJ....146...46K,2015ApJ...808L..27W}, in stark contrast to isolated field galaxies ($\mstar/\msun\in [10^7, 10^9]$) that are almost all actively star forming~\citep{geha2012, 2014ApJ...792..141S}. The majority of the detected LG low-mass galaxies are classical ($\mstar/\msun \in [10^5, 10^7]$) and bright ($\mstar/\msun \in [10^7, 10^9]$) dwarfs that show evidence of being quenched environmentally upon their accretion into the MW~\citep{2015ApJ...804..136W,2016MNRAS.463.1916F}. The majority of quenched LG satellites are also gas-poor~\citep{2009ApJ...696..385G,2014ApJ...795L...5S,2015ApJ...808L..27W,2021ApJ...913...53P}, suggesting tidal or ram-pressure stripping as potential quenching mechanisms. At even lower masses in UFDs, there is evidence for synchronous reionization quenching (e.g., \citealt{1996ApJ...465..608T,2000ApJ...539..517B,2002MNRAS.333..156B,2002MNRAS.333..177B,2012ApJ...759L..38A}) before they even entered either the MW or M31 and become impacted by environment (e.g., \citealt{2014ApJ...796...91B,2014ApJ...789..147W,2014ApJ...789..148W}).

Even though most satellites are quenched in the LG, the quenching timescales for MW and M31 satellites are significantly different. There exists a bimodal distribution of quenching timescales (one peak at 2 Gyrs ago and another peak at 10 Gyrs ago) for the MW satellites and a normal distribution (peaked at 6 Gyrs ago) for M31 satellites~\citep{2019ApJ...885L...8W,2021MNRAS.504.5270D}. This hints that host properties such as mass, environment, and assembly may significantly affect the star formation histories (SFHs) of low-mass galaxies. 

To probe dwarf satellite galaxies in broader ranges of environments beyond the LG ($\sim 5-100\,\mathrm{Mpc}$), several dedicated surveys have been conducted in the past decade. One such survey is the Satellites Around Galactic Analogs (SAGA) Survey \citep{Geha2017,Mao2021}, which characterizes satellite systems around MW-mass galaxies in 25-40.7\,Mpc.
Most recently, the SAGA Survey has published the third data release (DR3), which includes the census of 101 MW-mass satellite systems (\citealt{saga3:Mao2024}, hereafter \paperthree{}; \citealt{saga4:Geha2024}, hereafter \paperfour{}). This is the primary new dataset we use to constrain the model herein.

In addition to the SAGA Survey, the Exploring Local VolumE Satellites (ELVES, \citealt{2022ApJ...927...44C}) Survey, the ACS Nearby Galaxy Survey (ANGST, \citealt{2009ApJS..183...67D}), and the Dragonfly Wide Survey~\citep{2017ApJ...837..136D,2020ApJ...894..119D} have made significant recent progress in identifying and characterizing low-mass galaxies, especially in MW-mass host environments. These surveys have helped fill the gap between the nearby LG dwarfs that probe the extreme low-mass end of UFDs with the broader set of bright dwarfs from the Sloan Digital Sky Survey (SDSS, \citealt{2000AJ....120.1579Y,2004ApJ...613..898T,2007ApJS..173..267S}) at $z\gtrsim 0.1$. These newer low-mass galaxies populate a wide range of environments, extending the well-studied SDSS host-mass-dependent satellite quenched fractions~\citep{2012MNRAS.424..232W,2014MNRAS.442.1363W}, quenching time scales~\citep{2013MNRAS.432..336W,2014MNRAS.442.1396W,2016MNRAS.463.3083O}, and conformity with host galaxy star formation~\citep{2014MNRAS.437.1930P,2015MNRAS.447..698P}. Along with efforts focusing on local and more distant isolated field galaxies (e.g., \citealt{2023arXiv230519310L,2023ApJ...954..149D}) which are predominantly star-forming~\citep{geha2012,2021ApJ...913...45O}, studies of the low-redshift universe are flourishing due to growing data sets of 
low-mass galaxies in diverse cosmic environments. 

Theoretical efforts in modeling low-mass galaxy formation in connection to their dark matter halo assembly have embarked on various routes (see \citealt{2015ARA&A..53...51S} for a general overview of galaxy formation models and
\citealt{2018ARA&A..56..435W}  for an overview of galaxy--halo connection methodology). Modeling efforts using hydrodynamic simulations (see \citealt{2020NatRP...2...42V} for a review) have often focused on reproducing MW/M31-like dwarf satellite populations~\citep{2014MNRAS.438.2578G,2014MNRAS.445..581H,2017MNRAS.467..179G,2018MNRAS.480..800H,2018MNRAS.478..548S,2021ApJ...909..139A,2021MNRAS.508.1652J,2022MNRAS.511.1544F,2022MNRAS.514.5276S,2023MNRAS.522.5946E}. 
Other galaxy--halo modeling approaches such as semi-analytic models (SAMs; e.g., \citealt{1999MNRAS.310.1087S,2012NewA...17..175B,2021MNRAS.502..621J,2022MNRAS.514.2667K,2022MNRAS.516.3944M,2024MNRAS.529.3387A}), which are often combined with gravity-only zoom-in simulations, have successfully reproduced the properties of low-mass LG galaxies while keeping their evaluation costs lightweight relative to hydrodynamic simulations. Extensions of empirical models such as the subhalo abundance matching (SHAM) technique are even less complex and have yielded fruitful and statistically rigorous constraints on the stellar mass--halo mass (SMHM) relation of low-mass galaxies by leveraging the full MW satellite population~\citep{2019ApJ...873...34N,2020ApJ...893...48N}.

Despite this progress, both hydrodynamic simulations and SAMs have typically been too complex to have their full parameter space constrained by observations in a fully data-driven manner. To calibrate these physics-driven models against observational constraints, \citet{2023ApJ...949...94G} used the \textsc{SatGen}~\citep{2021MNRAS.502..621J} SAM to derive quenching timescales for ELVES satellites~\citep{2022ApJ...927...44C} for which they have re-tuned the SAM by hand to match observations. This is a significant step towards constraining SAM motivated by empirical observational data, although more work is needed to better understand the full posterior parameter space allowed by such models. Meanwhile, traditional abundance matching methods, which have significantly fewer parameters and are thus easier to constrain, only model the galaxy--halo connection at individual time stamps and do not self-consistently link galaxy evolution over time.

In this work, we present the first statistically constrained empirical galaxy--halo connection model based on a large sample of low-mass galaxies among a wide range of MW-mass hosts from the SAGA Survey (\paperthree{}) and an SDSS isolated field sample ~\citep{geha2012}. We aim to quantify with these new data the extent to which dark matter halo assembly accounts for the distribution of star formation and quenching in low-mass galaxies in a cosmological context.  In particular, we constrain the successful empirical galaxy--halo model \textsc{UniverseMachine}~(\citealt{2019MNRAS.488.3143B}) for the first time in the low-mass galaxy regime ($\mstar = 10^7 \sim 10^9 \msun$). This flexible empirical framework parameterizes galaxy SFRs with halo mass and assembly history, simultaneously matching observations over a wide range of galaxy masses and redshifts with a self-consistently time-evolving mock galaxy catalog anchored on simulated halo merger trees. Its flexible model for capturing galaxy SFR--halo assembly correlation guarantees that neither field- nor satellite-specific quenching mechanisms are imposed.

Previously, UM has been applied to cosmological~\citep{2020MNRAS.499.5702B} simulations at high redshift and to zoom-in~\citep{2021ApJ...915..116W} simulations with higher resolution that modeled galaxies down to $\mstar\sim 10^5 \msun$. Still, those extensions were model extrapolations and did not involve parameter re-calibration. Here, we add a new low-mass quenching model to UM in this work that enables low-mass galaxy quenching, which was a major limitation in the original UM model (hereafter `UM DR1'; see \citealt{2021ApJ...915..116W}). We constrain low-mass galaxy formation in this new UM model, `UM-SAGA', by jointly fitting to the newly added low-mass galaxy constraints from SAGA and SDSS at $z\sim0$, as well as to the comprehensive set of existing UM DR1 observations at higher masses covering a vast span of cosmic history ($z \sim 0-8$).

\begin{figure*}
    \centering
	\includegraphics[width=1.85\columnwidth]{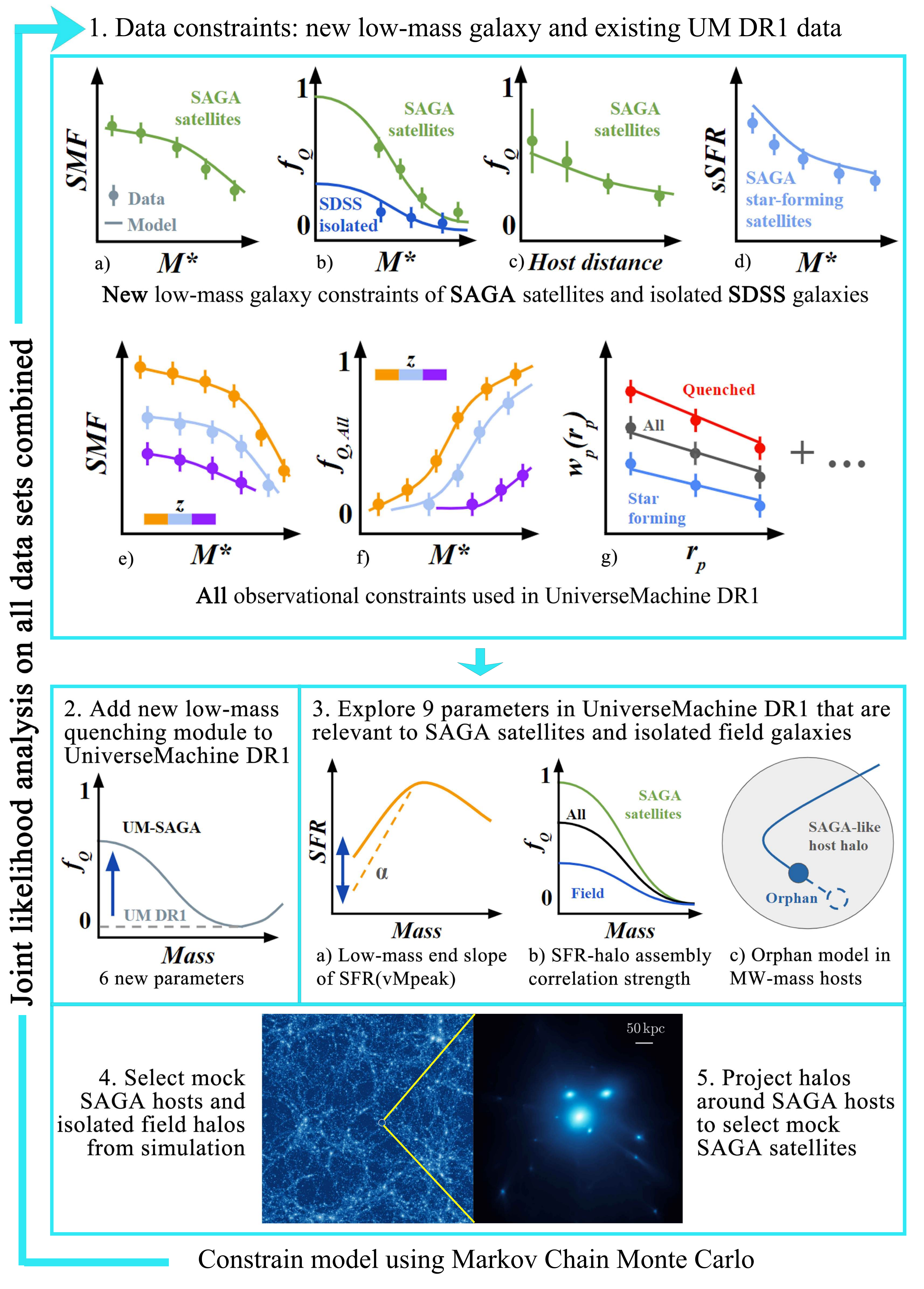}
    \caption{Schematic diagram of our modeling approach. The upper half highlights data constraints for UM-SAGA, consisting of newly added low-mass galaxy data (Section~\ref{sec:data}) and all other observations used in UM DR1. The lower half highlights model modifications to UM DR1 (Section~\ref{sec:model}) and how we select mock galaxy catalogs from the simulation (Section~\ref{sec:analysis}). SMF: stellar mass function; $w_p$: projected two-point correlation function; $r_p$: projected distance.}
    \label{fig:model}
\end{figure*}

The main advantage of this modeling approach is its ability to place low-mass galaxies in a broader cosmological context with their host and large-scale environment while constraining the underlying galaxy--halo connection with data. The predictions of this observationally constrained model will also be directly comparable to the low-mass galaxy formation histories in a wide range of hydrodynamic simulations and SAMs. Given the mass range ($\mstar \gtrsim 10^7 \msun$) of low-mass galaxy data employed in this work, we do not need to model the impact of reionization quenching \citep{2019arXiv190706653W}. Nonetheless, understanding the extent to which low-redshift environmental quenching in classical dwarfs can match observational data is crucial to understanding if and how galaxy quenching transitions to being dominated by reionization quenching, eventually leading to a unified framework for modeling star formation histories over the full range of halo masses that host observable galaxies (including $\mstar\leqslant 10^5\msun$ UFDs). In Fig.~\ref{fig:model}, we show a schematic diagram highlighting our methodology, including adopted observational data constraints, sample selection strategy, galaxy--halo modeling assumptions, and how the parameter space allowed by the new data is explored. 

The structure of the paper is as follows: in Section~\ref{sec:data}, we introduce the new low-mass galaxy constraints added from SAGA and SDSS to constrain the new UM model; in Section~\ref{sec:model}, we describe the new low-mass quenching model added to UM and the parts of the existing UM DR1 model that are relevant for low-mass galaxy formation and jointly constrained in this work; in Section~\ref{sec:analysis}, we describe how we select (sub)halos that host SAGA-like satellites and SDSS-like isolated field low-mass galaxies from the cosmological simulation Chinchilla c125-2048, on which we calibrate the UM-SAGA model; in Section~\ref{sec:results}, we present the new best-fit UM-SAGA model and how it compares to the input low-mass galaxy constraints as well as to the original UM DR1 model; in Section~\ref{sec:insights}, we discuss physical interpretations of the modeling results; in Section~\ref{sec:discussion} and potential limitations of the current model; in Section~\ref{sec:summary}, we summarize our conclusions and present a road map for future UM model upgrades. The best-fit UM-SAGA model and source code repository are publicly available at \url{https://bitbucket.org/RW-Stanford/universemachine-saga/src/main/}. 

\section{Data constraints for UM-SAGA}
\label{sec:data}

We aim to constrain UM-SAGA in the low-mass galaxy regime using both dwarfs found as satellites around MW-mass host galaxies (SAGA) as well as isolated field galaxies (SDSS) that sample distinct environments. This is a crucial upgrade from UM DR1 because it extends the mass range that the model is calibrated on and explicitly selects dwarfs in different environments, thus constraining the environmental quenching of low-mass galaxies.

We summarize the new low-mass galaxy constraints added to UM-SAGA in this work (Fig.~\ref{fig:model} panel 1 a, b, c, d):
\begin{enumerate}
    \item Stellar mass function of SAGA satellites ($\mstar>10^{7.5}\msun$, panel 1a);
    \item Quenched fraction as a function of stellar mass for SDSS isolated field galaxies ($\mstar>10^{7}\msun$, panel 1b);
    \item Quenched fraction as a function of stellar mass for SAGA satellites ($\mstar>10^{7.5}\msun$, panel 1b)
    \item Quenched fraction as a function of projected distance to host for SAGA satellites ($\mstar>10^{7.5}\msun$, $R_{\rm host}<300$ kpc, panel 1c); and
    \item Average NUV-specific SFR as a function of stellar mass for \emph{star-forming} SAGA satellites ($\mstar>10^{7.5}\msun$, panel 1d).
\end{enumerate}

We refer the reader to Papers III and IV for complete descriptions of these data products, as described in the following subsections. All SAGA satellite constraints are corrected for spectroscopic completeness. All the new low-mass galaxy constraints are at $z\lesssim 0.05$. These newly added data are schematically summarized in Fig.~\ref{fig:model}, and their actual values are shown in Fig.~\ref{fig:fqm}. We describe the SAGA and SDSS datasets in more detail below.

\subsection{SAGA Survey DR3 satellites}
\label{sec:data:saga}

We use data from the SAGA Survey~(\citealt{Geha2017, Mao2021}; \paperthree{}, \paperfour{}), which is a spectroscopic survey that targets faint galaxies around 101 MW-mass hosts from 25 to 40.75 Mpc. The SAGA DR3 sample includes 378 spectroscopically confirmed satellites within 300 kpc and line-of-sight velocity difference of $<275\ \mathrm{km\,s^{-1}}$ to their hosts. It is $94\%$ complete down $\mstar\geqslant 10^{7.5}\msun$, which is defined as the {\em Gold sample} (\paperthree{}). 

We use the Gold sample satellite stellar mass function (SMF, \paperthree{} Fig.~9 and Table~C.5), average satellite quenched fraction ($f_Q$) as a function of stellar mass (\paperthree{} Fig.~11 and Table~C.6; \paperfour{} Fig.~4 and Table~A.1), and projected distance to host (\paperfour{} Fig.~5 and Table~A.2), as well as average NUV-specific SFRs for star-forming satellites as a function of stellar mass (\paperfour{} Fig.~8 and Table~A.1) as data constraints to UM-SAGA. All SAGA constraints are corrected for spectroscopic incompleteness, except for the average NUV sSFR. We note that we use the \emph{combined} satellite SMF from all 101 hosts (i.e., we multiply the SMF per host shown in \paperthree{} by 101); we use four ([50, 100, 200, 300] kpc) instead of six radial bins for the $f_Q$ versus projected distance data shown in Table~A.2 of \paperfour{}, given that $f_Q$ is nearly constant beyond 100 kpc. The less complete {\em Silver sample} with $\mstar\geqslant 10^{6.75}\msun$ is not used to constrain the model, but we also show it in the following analysis for qualitative comparisons to the extrapolations of UM-SAGA.

\subsection{SDSS isolated field galaxies}
\label{sec:data:sdss}

The quenched fraction of isolated field galaxies with masses similar to the SAGA satellites is crucial information necessary to constrain satellite quenching. The $f_{Q}$ of isolated field galaxies sets the reference quenching intensity for low-mass galaxies at a certain mass scale in the absence of the influence from a larger host, capturing the efficiency of internal quenching processes such as stellar feedback~\citep[e.g.,][]{2014MNRAS.445..581H} and starvation~\citep[e.g.,][]{2017MNRAS.466.3460V} relative to environmental effects such as tidal and ram-pressure stripping.

We use the $f_Q(\mstar)$ data for isolated field low-mass galaxies in SDSS~\citep[Fig.~5]{geha2012} as another constraint on UM-SAGA. The isolated field sample (denoted Geha12 hereafter) consists of isolated low-mass galaxies that are the brightest objects within 1.5 Mpc (projected) of their vicinity out to $z< 0.055$ from the NASA Sloan Atlas catalog. The mass range of this sample spans $\mstar\geqslant 10^7 \msun$ to $\mstar\sim 10^{10}\msun$, which overlaps nicely with the SAGA DR3 satellite mass range. Since most galaxies in Geha12 are star-forming, it effectively adds a $f_Q (R_{\rm host} > 1.5\mathrm{Mpc}) \approx 0$ point to our radial quenching constraint.

\subsection{Existing data constraints in UM DR1} 
\label{sec:data:um-dr1}

The observational data constraints in UM DR1 (see Appendix of \citealt{2019MNRAS.488.3143B} for a full description) cover a wide range of masses ($\mstar \gtrsim 10^9\msun$) and redshifts ($z<8$). The types of data constraints include stellar mass functions, overall quenched fractions $f_{Q, \rm All}$ as a function of stellar mass, cosmic star formation rate density history, average specific SFR as a function of stellar mass, UV luminosity functions, infrared excess, the fraction of star-forming neighbors within 4 Mpc, as well as projected auto and cross-correlation functions for massive galaxies. These data are all inherited and \emph{all used} as constraints for UM-SAGA. Appendix~\ref{sec:App_B} shows detailed comparisons between how well UM-SAGA and UM DR1 match the input data constraints.

Some data constraints in UM DR1 are especially relevant for galaxy formation physics in the low-mass regime covered in this work, and these already constrained UM DR1 at $z\lesssim 0.1$ and $\mstar\lesssim 10^{8.5}\msun$. These constraints include (Fig.~\ref{fig:model} panel 1 e, f, g):
\begin{enumerate}
    \item GAMA Survey stellar mass functions~\citep{2012MNRAS.421..621B} at $z<0.06$ down to $\mstar\gtrsim 10^7\msun$ (panel 1e).
    \item GAMA Survey~\citep{2013MNRAS.434..209B} quenched fractions versus stellar mass for all galaxies down to $\mstar\gtrsim 10^9\msun$ (panel 1f).
    \item Projected two-point auto and cross-correlation functions from SDSS~\citep{2009ApJS..182..543A} for massive galaxies ($\mstar\geqslant 10^{10.3}\msun$) at $z<0.7$ (panel 1g).
\end{enumerate}

These low redshift constraints already used in UM DR1 have important implications when interpreting UM-SAGA results. In the following sections, we will discuss how these existing data connect to the newly added SAGA and SDSS low-mass galaxy data in more detail.

\section{The UniverseMachine Model}
\label{sec:model}

Here we briefly overview the modeling assumptions of UM DR1~\citep{2019MNRAS.488.3143B} and discuss the motivations for adding the new low-mass quenching model to create UM-SAGA. We also elaborate on the relative importance of existing parameters in UM DR1 in governing low-mass galaxy star formation predictions in UM-SAGA. We emphasize that the model update from UM DR1 to UM-SAGA does not change the fundamental model assumption of galaxy SFR being dependent only on halo mass and assembly history.

    Section~\ref{sec:model:um-dr1} summarizes the key model assumptions of UM DR1. It shows how galaxy SFR is generated based on halo mass ($v_{\rm Mpeak}$) and assembly history ($\Delta v_{\rm max}$), with a special focus on the physical insights of the SFR--$\Delta v_{\rm max}$ correlation through the parameter $r_c$.
    Section~\ref{sec:model:fq} introduces the new low-mass quenching model components (Fig.~\ref{fig:model} panel 2) added in UM-SAGA to address the limitation of UM DR1 in the low-mass galaxy regime.
    Section~\ref{sec:model:dr1-free} describes the relevant low-mass galaxy parameters in UM DR1 that are explored and jointly constrained by the new low-mass galaxy data. re-constrain the SFR--halo mass relation low-mass end slope (Fig.~\ref{fig:model} panel 3a), the SFR--halo assembly rank correlation strength (Fig.~\ref{fig:model} panel 3b), and the orphan tracking model around MW-mass hosts (Fig.~\ref{fig:model} panel 3c) . 

\subsection{Summary of the UM DR1 model}
\label{sec:model:um-dr1}

We begin with an overview of the UM DR1 model that UM-SAGA builds on. The empirical galaxy--halo connection model predicts a galaxy's SFR given its (sub)halos' halo mass (parameterized by the maximum circular velocity at peak historic halo mass $v_{\mathrm{Mpeak}}$), halo assembly history ($\Delta v_{\rm max}$, change in $v_{\mathrm{max}}$ over time), and redshift, i.e., $P(\mathrm{SFR}| v_{\rm Mpeak}, \Delta v_{\rm max}, z)$. We use $v_{\rm peak}$ as a proxy for halo mass since it unifies both host and subhalos to a `pre-infall' state. This definition also has the advantage that it is immune to cosmological pseudo-evolution~\citep{2006ApJ...647..201C,2013ApJ...766...25D} that occurs for the overdensity-based halo mass definition. The model paints SFRs onto halo merger trees from cosmological gravity-only simulations and self-consistently models the cosmic evolution and assembly of galaxies by integrating the predicted SFRs along halo merger trees. The model is then constrained by galaxy observations across cosmic history (at $0<z<8$), including stellar mass/UV luminosity functions, quenched fractions, cosmic star formation rate density, and two-point correlation functions.

More specifically, we begin by parameterizing the fraction of quenched galaxies ($f_{Q}$) in the universe as a function of halo mass and redshift $f_Q(v_{\rm Mpeak}, z)$ -- we will return to the exact parameterization of the quenched fraction and the low-mass quenching model introduced by this work in \autoref{sec:model:fq}. At a given redshift and halo mass scale, UM describes the probability density distribution (PDF) of SFRs for (sub)halos in the $p(\mathrm{SFR}|v_{\mathrm{Mpeak}, i}$, $z_j)$ bin as a double-Gaussian bimodal distribution,  wherein the two peaks of the PDF correspond to the quenched and star-forming populations. The average SFR of the star-forming population is modeled as a redshift-dependent double power-law function of halo mass SFR$(v_{\rm Mpeak}, z)$. The average SFR of the quenched population is set to a fixed specific SFR of $\log_{10} (\mathrm{sSFR}_Q/\rm yr^{-1}) = -11.8$, motivated by SDSS observations~\citep{2015MNRAS.450.1546B}. The normalization ratio of the Gaussian peaks in each $(v_{\mathrm{Mpeak}, i}, z_j)$ bin is given by $f_Q/(1-f_Q)$; the scatter for each Gaussian is also redshift dependent.

To assign the SFR of individual (sub)halos in a given ($v_{\rm Mpeak, i}$, $z_j$) bin using this probabilistic $p(\mathrm{SFR}|v_{\mathrm{Mpeak}, i}$, $z_j)$ model, 
UM rank-correlates the assembly history parameter $\Delta v_{\rm max}$ of each (sub)halo to their SFRs. The halo assembly parameter $\Delta v_{\rm max}$ is defined as:
\begin{equation}
\label{eq:3}
    \Delta v_{\rm max}(z) = \frac{v_{\rm max} (z)}{v_{\rm max}(\max \left[z_{\rm dyn}, z_{\rm Mpeak}\right])} ,
\end{equation}
where $z_{\rm dyn}$ is the redshift one dynamical timescale ago from $z$ and $z_{\rm Mpeak}$ is the redshift when the (sub)halo reached its peak halo mass. The advantage of using $\Delta v_{\rm max}$ over a discrete classification of halos into centrals and subhalos is its continuous depiction of halo assembly histories that can be mapped onto a continuous distribution of SFRs. It was chosen over alternative parameterizations of the halo assembly history such as infall times, mass accretion history, and tidal forces because it is less prone to systematic uncertainties in simulations and better physical intuition (see Appendix A in \citealt{2019MNRAS.488.3143B} for more details). This allows us to model satellite and ``isolated'' galaxy quenching on the same footing without imposing an artificial boundary (e.g., $R_{\rm host}$) around host halos where satellite quenching begins to occur in an abrupt manner. Intuitively, for halos with the same mass, halos growing over the past dynamical timescale (mainly in the field) have $\Delta v_{\rm max}>1$ and tend to be assigned a higher SFR, while halos being stripped (mostly satellite subhalos) over the past dynamical timescale have $\Delta v_{\rm max}<1$ and tend to be assigned a lower SFR. The rank-correlation strength between SFR and $\Delta v_{\rm max}$ is a free parameter in UM that {\em characterizes the correlation strength between halo assembly and galaxy quenching.} This parameter is mainly constrained by galaxy clustering observations (auto and cross-correlation functions) from SDSS at $\mstar \gtrsim 10^{10}\msun$ in UM DR1 and quantifies how strongly star formation is tied to halo growth.

With the aforementioned $P(\mathrm{SFR}| v_{\rm Mpeak}, \Delta v_{\rm max}, z)$ model, UM predicts an SFR for each (sub)halo in the simulation at every snapshot. This can then be integrated along the halo merger trees to model (with appropriate observational systematics applied) a time-evolving galaxy catalog that naturally includes in-situ star formation and ex-situ stellar mass acquired from mergers. The self-consistent time-evolution aspect of UM is one of its primary advantages over traditional single-redshift abundance matching methods, as it allows UM to simultaneously model galaxy evolution in observational datasets that span a wide range of masses and redshift epochs in a causally correlated manner. UM DR1 also includes modules for stochasticity in short-timescale SFR fluctuations, stellar mass loss due to dying massive stars, an intra-cluster-light (ICL) component for each halo, and, most importantly, for the present work, an orphan model for tracing galaxies after their subhalos disrupt.

As mentioned in Section~\ref{sec:intro}, UM DR1 directly applied to high-resolution zoom-in simulations~\citep{2021ApJ...915..116W} showed an SMHM relation that is reasonably consistent with the latest abundance matching constraints from MW satellite observations~\citep{2020ApJ...893...48N}. However, this concordance was reached by requiring low-mass galaxies at $\mstar \lesssim 10^8 \msun$ to form a significant proportion of their stars at $z\lesssim 0.5$, which is inconsistent with observationally derived low-mass galaxy SFHs in the Local Group~\citep{2014ApJ...789..147W}. The reason why UM DR1 creates this late star formation artifact is because of extrapolating the error function parameterization of $f_Q(v_{\rm Mpeak})$ (see the first line of Eq.~\ref{eq:4}) into the low-mass galaxy regime. This extrapolation enforces all low-mass galaxies with $ 10^7\msun  \lesssim \mstar \lesssim 10^8\msun$ to be star-forming ($f_Q \longrightarrow 0$) throughout cosmic history. Although isolated low-mass galaxies seem to be consistent with $f_Q \approx 0$ (Section~\ref{sec:data:sdss}, \citealt{geha2012}), it is clear that low-mass satellites around MW-mass hosts~(Section~\ref{sec:data:saga}, \citealt{2015ApJ...808L..27W,Mao2021}; \paperthree{}) have a significant quenched population. This inflexibility motivates a new model for low-mass galaxy quenching.

\subsection{The new low-mass quenching model}
\label{sec:model:fq}

To allow a nonzero quenched fraction for (sub)halos that host low-mass galaxies at $\mstar \lesssim 10^9\msun$, we extend the parameterization of $f_Q(v_{\rm Mpeak})$ in UM to lower masses by adding in another set of independent error functions that allow for $f_Q > 0$. Inspired by the UM DR1 parameterization of $f_{Q}(v_{\rm Mpeak})$ we parameterize the \emph{overall} quenched fraction of galaxies as a function of their (sub)halo $v_{\rm Mpeak}$ and redshift $z$ via
\begin{equation}
\label{eq:4}
\begin{aligned}
f_{Q} & =Q_{\min}+\left(1-Q_{\min}\right){\left[0.5+0.5 \operatorname{erf}\left(\frac{\log _{10}\left(v_{\mathrm{Mpeak}} / v_{Q1}\right)}{\sqrt{2} \sigma_{VQ1}}\right)\right] } \\
& +\left(1-Q_{\min}\right)\left[0.5-0.5 \operatorname{erf}\left(\frac{\log _{10}\left(v_{\mathrm{Mpeak}} / v_{Q2}\right)}{\sqrt{2} \sigma_{VQ2}}\right)\right],
\end{aligned}
\end{equation}
in which the model parameters $Q_{\rm min}$, $v_{Q1}$, $\sigma_{VQ1}$, $v_{Q2}$, $\sigma_{VQ2}$ are all redshift dependent. The first two terms in Eq.~\ref{eq:4} are simply the UM DR1 $f_Q(v_{\rm Mpeak})$ model. These account for quenching at higher galaxy masses (constrained to $\mstar \gtrsim 10^9\msun$). The parameters that govern the redshift evolution of $Q_{\rm min}$, $v_{Q1}$, and $\sigma_{VQ1}$ are fixed to their UM DR1 best-fit values.

The last term in Equation~\ref{eq:4} is the {\em new low-mass quenching model} that allows for non-monotonic \emph{overall} quenched fractions at lower masses galaxy and halo masses. This new module allows $f_Q$ to rise again towards lower masses ($\mstar\lesssim 10^9\msun$) instead of asymptoting to zero as in UM DR1. There are six new parameters added to the quenching model; these are coefficients for the redshift dependence of $v_{Q2}$ and $\sigma_{VQ2}$:
\begin{equation}
\label{eq:5}
    \begin{aligned}
\log _{10}\left(V_{Q2}\right) & =V_{Q2, 0}+V_{Q2, a}(1-a)+V_{Q2, z} z \\
\sigma_{V Q2} & =\sigma_{VQ2, 0}+\sigma_{VQ2, a}(1-a)+\sigma_{VQ2, l a} \ln (1+z)\,,
\end{aligned}
\end{equation}
which are similar to the redshift parameterization for $v_{Q1}$ and $\sigma_{VQ1}$. We also enforce a ceiling value for $f_Q \leqslant 1$. We show in the second row of Fig.~\ref{fig:model} the schematic effect of adding the new low-mass quenching module to UM DR1. To better visualize the effects of the new low-mass quenching model, we show the best-fit UM-SAGA $f_Q(v_{\rm Mpeak})$ functions compared to the UM DR1 counterpart in Appendix~\ref{sec:App_B} (Fig.~\ref{fig:B0}). Furthermore, we demonstrate in the Appendix~\ref{sec:App_A} (Fig.~\ref{fig:A3}) that the new dwarf-galaxy quenching module has a negligible correlation with existing modules in UM DR1 that affect low-mass galaxies, truly empowering UM-SAGA with crucial degrees of freedom that the DR1 model was not capable of.

\subsection{Existing UM DR1 parameters relevant to low-mass galaxy star formation}
\label{sec:model:dr1-free}

\begin{table*}
		\centering
		\begin{tabular}{lccc}
			\hline
            \hline
			Parameter & Physical meaning & UM-SAGA $68\%$ Posterior & UM DR1 $68\%$ posterior\\
			\hline
            $\alpha_{0}$ & $\mathrm{SFR}(v_{\rm Mpeak})$ low-mass end slope at $z=0$ & $-6.14^{+0.26}_{-0.27}$  & $-5.97^{+0.59}_{-0.62}$\\
            $\alpha_{a}$ & Evolution of $\alpha$ at low redshift & $-3.93^{+0.90}_{-0.90}$ & $-6.39^{+3.05}_{-2.69}$\\
            $\alpha_{z}$ & Evolution of $\alpha$ at high redshift & $-0.54^{+0.15}_{-0.16}$ & $-1.01^{+0.64}_{-0.55}$\\
            $\alpha_{la}$ & Evolution of $\alpha$ at medium redshift & $6.37^{+1.24}_{-1.21}$ & $10.64^{+4.02}_{-5.07}$\\
            $r_{\min}$ & Minimum $\mathrm{SFR} \sim \Delta v_{\rm max}$ correlation & $0.48^{+0.02}_{-0.02}$  & $0.10^{+0.18}_{-0.19}$\\
            $r_{\rm width}$ & Width (slope) of $r_c(v_{\rm Mpeak})$ & $0.19^{+0.07}_{-0.07}$ & $5.96^{+2.90}_{-3.24}$\\
            $V_{R, 0}$ & Halo $v_{\rm Mpeak}$ scale where $r_c = (1+r_{\rm min})/2$ & $2.24^{+0.07}_{-0.06}$ & $2.27^{+1.19}_{-1.52}$\\
            $V_{R, a}$ & Linear redshift evolution of $V_{R, 0}$ & $-5.72^{+1.92}_{-1.55}$ & $-1.78^{+1.94}_{-2.55}$\\
            $T_{\rm merge, 300}$ & Orphan tracking threshold $(v_{\rm max}/v_{\rm Mpeak})$ in hosts of $v_{\rm Mpeak}=300\,\mathrm{km\,s^{-1}}$  & $0.71^{+0.01}_{-0.01}$   & $0.57^{+0.05}_{-0.07}$\\
            \hline
			$V_{Q2, 0}$ & Halo $v_{\rm Mpeak}$ where $f_Q=0.5$ at $z=0$ & $1.66^{+0.01}_{-0.02}$ & - \\
			$V_{Q2, a}$ & Low-redshift evolution of the $f_Q=0.5$ $v_{\rm Mpeak}$ scale & $-0.23^{+0.18}_{-0.47}$  & - \\
			$V_{Q2, z}$ & High-redshift evolution of the $f_Q=0.5$ $v_{\rm Mpeak}$ scale & $-0.63^{+0.33}_{-0.32}$  & - \\
            $\sigma_{VQ2, 0}$ & Width of the $f_Q(v_{\rm Mpeak})$ error function at $z=0$ & $0.14^{+0.02}_{-0.01}$  & - \\
			$\sigma_{VQ2, a}$ & Low-redshift evolution of the $f_Q(v_{\rm Mpeak})$ width & $0.38^{+0.28}_{-0.34}$ & - \\
			$\sigma_{VQ2, z}$ & High-redshift evolution of the $f_Q(v_{\rm Mpeak})$ width & $0.22^{+0.69}_{-0.47}$ & - \\
			\hline
		\end{tabular}
		\caption{The physical meaning, posterior median, and $[16\%, 84\%]$ marginalized posterior distribution for the 15 parameters constrained by new low-mass galaxy observation data in UM-SAGA. The upper half consists of 9 parameters from UM DR1 relevant to low-mass galaxies that we explore; the lower half consists of six new parameters introduced in UM-SAGA that account for low-mass galaxy quenching. We also show the original $68\%$ posteriors in~\citet{2019MNRAS.488.3143B} for the 9 parameters from UM DR1.}
		\label{tab:1}
\end{table*}

With the addition of the six new low-mass quenching parameters, we would like to re-constrain the complete set of 52 parameters in UM with the existing DR1 constraints and new low-mass galaxy constraints (Section~\ref{sec:data}). However, the computational demand of exploring the full parameter space of UM is formidable. Since many of the original 46 UM DR1 parameters were well-constrained and only relevant at higher mass scales, re-constraining the bulk part of the model would be redundant. Furthermore,  even if we did have the resources to refit all 52 UM parameters on c125-2048, it would be far from ideal to model observables such as the two-point correlation functions using c125-2048 due to its relatively small cosmological volume. Therefore, we only explore parts of the UM DR1 model essential for modeling star formation in the low-mass galaxy regime that are best constrained by the new low-mass galaxy data added in this work. 

As an overview, we explore nine parameters from the UM DR1 model relevant to low-mass galaxy formation and keep the other 37 parameters from UM DR1, which are unimportant to low-mass galaxies, fixed. These nine parameters consist of four parameters for the low-mass end slope of the SFR--halo mass scaling, four for the rank correlation between SFR and halo assembly, and one for orphan tracking around MW-mass hosts. The physical meaning of these nine UM DR1 parameters along with the six new low-mass quenching parameters introduced in Section~\ref{sec:analysis:hosts} are summarized in Table~\ref{tab:1}.

\subsubsection{Low-mass slope of the $\mathrm{SFR} (v_{\rm Mpeak})$ scaling relation}
\label{sec:model:sfr-slope}
  
With the addition of the new low-mass quenching model in Section~\ref{sec:analysis:hosts}, low-mass galaxies at $\mstar \lesssim 10^8 \msun$ will start to quench earlier in their cosmic evolution, which immediately leads to a reduction in the  $z=0$ stellar masses of these dwarfs at fixed halo mass. As shown in \citet{2021ApJ...915..116W}, UM DR1 produces a SMHM relation that is consistent with the latest abundance matching constraints based on MW satellites, driven by the SMF constraints from the GAMA Survey~\citep{2012MNRAS.421..621B} used in UM DR1 down to $\mstar\sim 10^7\msun$. To maintain a similar SMHM relation at $z=0$ for low-mass galaxies, the model has to boost the SFR in low-mass galaxies to compensate for the added quenching by the new model. Hence, we explore  the redshift-evolving low-mass slope of the double-power-law $\mathrm{SFR} (v_{\rm Mpeak})$ scaling, where $\mathrm{SFR} (v_{\rm Mpeak}) \propto \left(v_{\rm Mpeak}^{\alpha(z)}+ v_{\rm Mpeak}^{\beta(z)}\right)$ and $\alpha(z)$ follows
\begin{equation}
\label{eq:6}
    \alpha(z) = \alpha_0 + \alpha_a(1-a) + \alpha_{la} \ln(1+z) + \alpha_{z} z,\ a = (1+z)^{-1}.
\end{equation}
We explore all 4 parameters $\alpha_0$, $\alpha_a$, $\alpha_{la}$, and $\alpha_z$ during MCMC analysis of the UM-SAGA model; see Fig.~\ref{fig:model}, panel 3a.

\subsubsection{Rank correlation coefficient between SFR and $\Delta v_{\rm max}$}
\label{sec:model:rc}

In a certain halo $v_{\rm Mpeak}$ bin, UM rank correlates halo assembly with galaxy SFR using the rank correlation coefficient $r_c$:
\begin{equation}
\label{eq:8}
    C({\rm SFR}) = C\left(r_c C^{-1}(C(\Delta v_{\rm max})) + \sqrt{1-r_c^2} R_{N} \right),
\end{equation}
where $C(x) = 0.5 + 0.5 {\rm erf} (x/\sqrt{2})$ (cumulative percentile rank function) and $R_N$ is a standard normal random variable that introduces scatter to the correlation. With the new low-mass quenching model and a re-constrained low-mass $\mathrm{SFR} (v_{\rm Mpeak})$ scaling, the UM-SAGA model would be able to produce realistic SFR distributions for low-mass galaxies. However, to faithfully paint SFRs onto individual halos, the model must also know the correlation strength between halo growth and star formation to capture the effect of environmental quenching. The correlation strength, $r_c$, which rank correlates SFR and halo accretion status $\Delta v_{\rm max}$, was quite poorly constrained in UM DR1 by a limited set of two-point correlation functions for star-forming and quenched galaxies at $\mstar>10^{10.3}\msun$ and $z< 0.7$. Since $r_c$ is only constrained for high-mass galaxies at relatively low redshift, the UM DR1 model is ignorant of the strength of the halo growth--star formation correlation for low-mass galaxies, and we expect the addition of SAGA satellite and SDSS isolated field low-mass galaxy data at $\mstar\lesssim 10^9\msun$ could significantly improve the constraints on $r_c$. Therefore, we also explore the $r_{c} (v_{\rm Mpeak})$ part in UM DR1:
\begin{equation}
\label{eq:7}
    \begin{aligned}
r_c(v_{\text {Mpeak}})= & r_{\min }+\left(1.0-r_{\min }\right) \times \\
& {\left[0.5-0.5 \operatorname{erf}\left(\frac{\log _{10}\left(v_{\text {Mpeak}}/V_R \right)}{\sqrt{2} \cdot r_{\text {width }}}\right)\right] },
\end{aligned}
\end{equation}
where the characteristic halo $v_{\rm Mpeak}$ scale is redshift dependent following $\log _{10}\left(V_R\right)=V_{R, 0}+V_{R, a}(1-a)$. We vary all four parameters related to $r_c$: $r_{\min}$, $r_{\rm width}$, $V_{R,0}$, and $V_{R, a}$, while constraining the UM-SAGA model; see Fig.~\ref{fig:model}, panel 3b.

\subsubsection{Orphan model}
\label{sec:model:orphan}
The orphan model in UM (not redshift-dependent) is a numerical model that accounts for galaxies that live in subhalos artificially lost by the halo finder due to severe tidal stripping (see Section 3.3 and Appendix B in \citealt{2019MNRAS.488.3143B}). The extent to which it is necessary depends on the numerical resolution of the simulation and the subsequent halo-finding method~\citep{2023arXiv230810926M}. For the resolution of our c125-2048 box and the halo finder used, the UM orphan model is crucial to model galaxy number densities faithfully. UM adopts various semi-analytic models to predict the spatial trajectories and mass-loss process of orphan galaxies following the disruption of their subhalos. The duration of tracking orphan galaxies affects the overall density of galaxies over time. Furthermore, since orphan galaxies originate from disrupted subhalos that are heavily stripped, they tend to be concentrated near host halo centers, which could significantly alter quenched fraction as a function of satellite mass and distance to the host. Therefore, the orphan model is another critical degree of freedom to consider when matching the new low-mass galaxy constraints from SAGA and SDSS. 

In UM DR1, the host-mass dependent orphan tracking threshold, $T_{\rm merge}$ is a function of host halo mass (parameterized by $v_{\rm Mpeak, host}/ \mathrm{km\,s^{-1}}$):
\begin{equation}
\label{eq:Tmerge}
\begin{aligned}
    T_{\rm merge} &= T_{\rm merge, 300} + (T_{\rm merge, 1000}-T_{\rm merge, 300}) \\
    & \quad \times \left(0.5 + 0.5\mathrm{erf}(\frac{\log_{10}v_{\rm Mpeak, host} - 2.75}{0.25\sqrt{2}}) \right).
\end{aligned}
\end{equation}
It varies smoothly between host mass $v_{\rm Mpeak, host}$ of 300 $\mathrm{km\,s^{-1}}$ and 1000 $\mathrm{km\,s^{-1}}$). Orphan satellites are tracked to the point when they are tidally stripped to $v_{\rm max}(z_{\rm now})/v_{\rm Mpeak} = T_{\rm merge}$ after falling into their hosts. We explore the $T_{\rm merge, 300}$ parameter in our analysis, which is the low-host-mass end orphan threshold for orphans around MW-mass hosts with $v_{\rm Mpeak, host} = 300\ \mathrm{km\,s^{-1}}$ which is relevant for SAGA-like host environments; see Fig.~\ref{fig:model}, panel 3c. We apply orphan tracking to (sub)halos with $v_{\rm max} > 35\ \mathrm{km\,s^{-1}}$ in this work which is equivalent to the (sub)halo resolution limit in c125-2048 (UM DR1 only applied to (sub)halos with $v_{\rm max} > 80\ \mathrm{km\,s^{-1}}$ given the coarser resolution of \emph{Bolshoi-Planck}). 

Finally, if a subhalo is stripped below $T_{\rm merge}$ but is still tracked by \textsc{consistent-trees}, UM still considers the galaxy disrupted. This effectively takes into account the effect of baryonic disruption due to the central galaxy's potential~\citep[e.g.,][]{2017MNRAS.471.1709G,2018ApJ...859..129N,2020ApJ...893...48N} in addition to subhalos artificially lost by the halo finder numerically. We expect that the combination of more robust subhalo tracking methods~\citep[e.g., \textsc{symfind}][]{2023arXiv230810926M} and dark matter zoom-in simulations with embedded analytic disk potentials~\citep{2024arXiv240801487W} will enable a more realistic model for subhalo disruption. This will more clearly disentangle the necessity of actual orphan satellites from an effective model that accounts for both orphans and disk disruption.

\section{Analysis setup}
\label{sec:analysis}

We now introduce the analysis procedure used to constrain UM-SAGA with newly added low-mass galaxy data and existing UM DR1 data. We briefly summarize the properties of the cosmological simulation c125-2048 from which we select SAGA-like hosts and subhalos and SDSS-like isolated field halos. These selected (sub)halo catalogs, which mimic SAGA and SDSS selection criteria, are fed into the model analysis pipeline for generating mock observables (SFR and $\mstar$, Section~\ref{sec:model}), which are then fitted to observations using Markov Chain Monte Carlo (MCMC).

Section~\ref{sec:analysis:sim} introduces dark matter halo catalogs and merger trees from the Chinchilla c125-2048 cosmological simulation on which we anchor our analysis (Fig.~\ref{fig:model}, panel 4).
Section~\ref{sec:analysis:hosts} summarizes how SAGA-like hosts are selected from these halo catalogs (Fig.~\ref{fig:model}, panel 4).
Section~\ref{sec:analysis:sats} summarizes how SAGA-like satellites living in both surviving and orphan subhalos are selected based on the mock SAGA-like hosts (Fig.~\ref{fig:model}, panel 5).
Section~\ref{sec:analysis:sdss} summarizes how SDSS isolated field galaxies are selected in the simulation (Fig.~\ref{fig:model}, panel 4).
Section~\ref{sec:analysis:priors} summarizes the parameter prior choices.
Section~\ref{sec:analysis:mcmc} summarizes the MCMC setup for constraining UM-SAGA.

\subsection{The Chinchilla c125-2048 simulation}
\label{sec:analysis:sim}

We use the cosmological box `Chinchilla' c125-2048~\citep{2015ApJ...810...21M} to fit the UM-SAGA model to the new low-mass galaxy observations. The c125-2048 box is a gravity-only simulation run with \textsc{L-Gadget} (a gravity-only version of \textsc{Gadget-2}, \citealt{2005MNRAS.364.1105S}) with a side length of 125 Mpc~$h^{-1}$ (using periodic boundary conditions), force softening scale $\epsilon = 0.71$ kpc, and mass resolution of $m_{\rm DM}=2.57 \times 10^7\msun$. 

Since our new low-mass galaxy constraints include galaxies with $\mstar\sim 10^7 \msun$ (Section~\ref{sec:data}), the simulation should resolve (sub)halos with peak halo masses $M_{\mathrm{peak}}\sim 2\times 10^{10}\msun$ according to the UM DR1 model SMHM relation. Therefore, the numerical resolution of c125-2048 is well-suited for this work. 
A detailed numerical convergence study of this box was presented in the appendix of \citet{2023ApJ...945..159N}. 

The c125-2048 simulation adopts a flat $\Lambda$CDM cosmological model with $h_{0} = 0.7$, $\Omega_{\mathrm{m}} = 0.286$, $\Omega_{\Lambda} = 0.714$, $\sigma_{8} = 0.82$, and $n_{\mathrm{s}} = 0.96$. Dark matter (sub)halos (assuming virial overdensity $\Delta_{c} (z=0) = 99.2$) are identified with the \textsc{Rockstar} phase-space halo finder while halo merger trees are planted using \textsc{Consistent-Trees}~\citep{2013ApJ...763...18B}. All comparisons to SAGA satellites are conducted using the $z=0$ snapshot of the simulation, while comparisons to SDSS isolated field galaxies cover three snapshots: $z=0, 0.025, 0.054$.

We note that UM DR1 was calibrated on a different cosmological simulation \emph{Bolshoi-Planck}~\citep{2016MNRAS.457.4340K,2016MNRAS.462..893R}, which is a 250 Mpc~$h^{-1}$ side length box with a mass resolution of $m_{\rm DM}\sim 2.2\times 10^9\msun$ and different flat $\Lambda$CDM cosmology ($h = 0.678$, $\Omega_{\rm m} = 0.308$). Due to the changes in box size, mass resolution, and cosmology, UM DR1 applied onto Chinchilla c125-2048 predicts a slightly different universe of galaxies compared to \emph{Bolshoi-Planck} with the same set of model parameters, especially for the stellar mass functions and two-point correlation functions due to simulating a fundamentally different patch of the Universe with a different cosmology. 

Since these changes are small compared to observational errors and our goal for this paper is to extend UM to the low-mass galaxy regime, we opt \emph{not to re-calibrate} UM DR1 on c125-2048 as a bulk part of the DR1 model will not be affected during the calibration process onto the new low-mass galaxy constraints. Instead, we apply UM DR1 onto both \emph{Bolshoi-Planck} and c125-2048 to retrieve model prediction differences for \emph{all observational} data points used in UM DR1. We shift all \emph{observational} data used in UM DR1 by the model prediction differences at each data point to mimic the change in cosmology, particle resolution, and box size going from \emph{Bolshoi-Planck} to c125-2048. This procedure also guarantees that the total $\chi^2$ of the UM DR1 model is kept fixed going from \emph{Bolshoi-Planck} to c125-2048. We include all the \emph{shifted} UM DR1 data constraints in the MCMC likelihood analysis of the UM-SAGA model and the newly added low-mass galaxy constraints in Section~\ref{sec:data}.

\subsection{SAGA-like host halo selection}
\label{sec:analysis:hosts}

The primary selection criteria for SAGA host galaxies is their $K$-band absolute magnitude, $-24.6<M_K<-23$~(\citealt{Mao2021}; \paperthree{}). However, at present, UM only predicts galaxy stellar masses, and one needs to assume a $K$-band stellar-mass-to-light ratio to select mock MW-mass galaxies as SAGA-like hosts based on $M_K$. To circumvent the potential bias caused by introducing a fixed stellar mass-to-light ratio, we follow the host selection method in SAGA DR1~\citep{Geha2017} and DR2~\citep{Mao2021} by performing abundance matching of halo peak maximum circular velocity ($v_{\rm peak}$) directly to host galaxy $M_K$. 

Specifically, halos in c125-2048 are abundance-matched to $M_K$ given their $v_{\rm peak}$ using the 6dF Galaxy Survey luminosity function~\citep{2006MNRAS.369...25J}. A scatter of 0.15 dex is assumed during abundance matching for the central luminosity. We define central halos with an abundance matched $-24.6 < M_{K, \rm host} < -23$ as `SAGA-like' hosts in the c125-2048 simulation. To mimic SAGA-like host isolation criteria~(see Section 2.1.2 in \citealt{Mao2021}), we further select hosts that are the heaviest halo within three times its own virial radius $3R_{\rm vir, host}$, and we require their brightest neighbor projected (along $z$-axis of the box) within 300 kpc of the host halo to be dimmer than a magnitude gap of $M_{K, \rm host} - M^{<300\rm\,kpc}_{K, \rm brightest} < -1.6$. We also exclude a small number of hosts with $M_{\rm vir} > 10^{13}\msun$ following \citet{Mao2021}. With the combination of the $M_K$ range and isolation criteria, we end up with a SAGA-like host halo catalog of 7473 halos in c125-2048. 

\begin{figure}[t!]
	\includegraphics[width=\columnwidth]{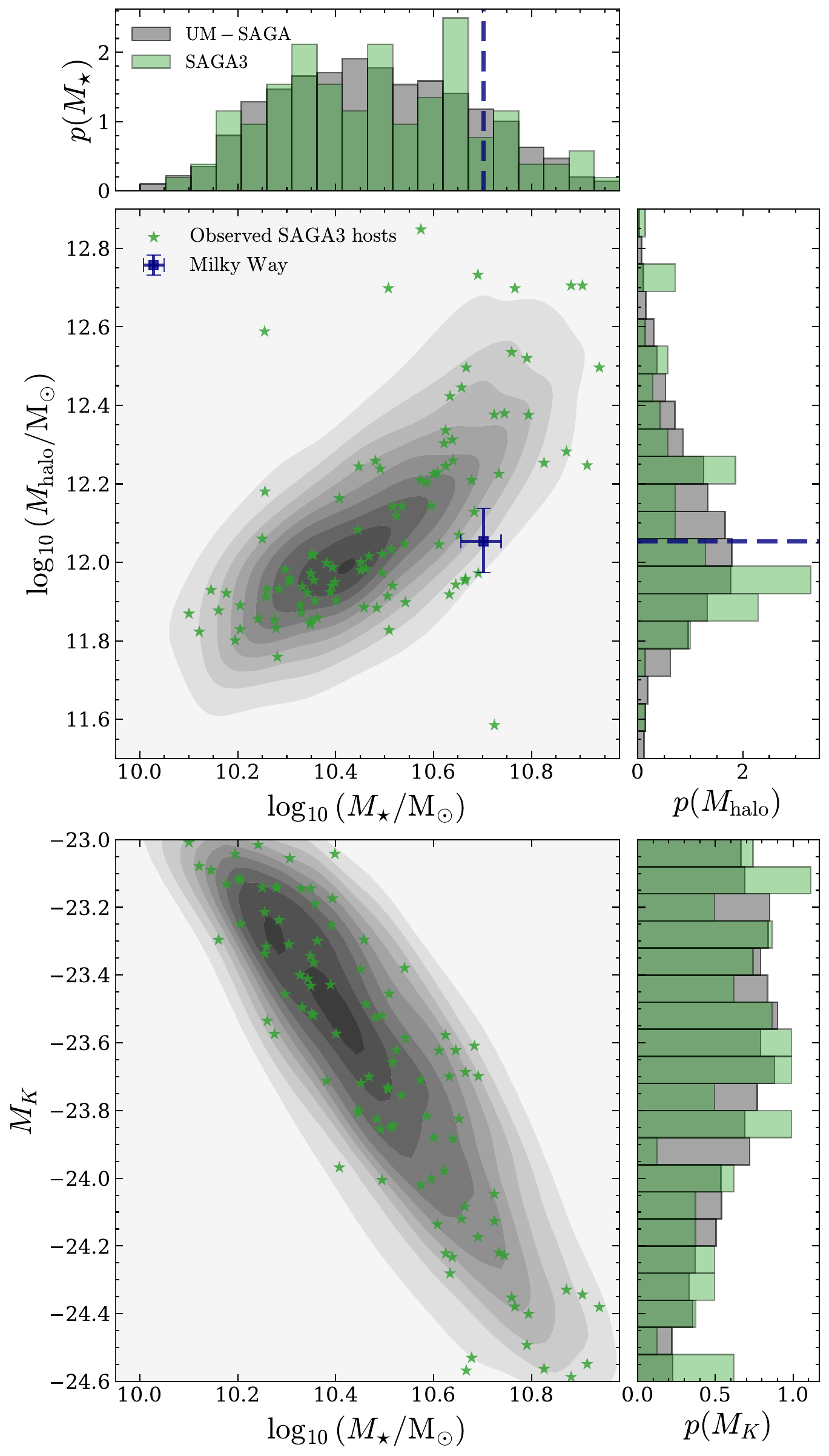}
    \caption{Distribution of stellar mass versus halo mass (upper panel) and $K$-band magnitude (lower panel) of 2500 mock SAGA-like hosts in c125-2048 (gray contours, $10\%$ intervals), which well-reproduce the distributions of the 101 observed SAGA hosts (red stars). The stellar mass values for the SAGA hosts are converted from their observed $K$-band magnitude (\paperthree{}); their halo masses are group masses cross-matched to the 2MRS catalog~\citep{2017MNRAS.470.2982L}. The Milky Way's observationally inferred stellar and halo masses are from \citet{2020MNRAS.494.4291C}. We convert the $M_{200}$ halo masses for the Milky Way to $\Delta_{c} (z=0) = 99.2$ virial masses using their observationally inferred halo concentration. We quote the UM-SAGA $\mstar$ values for these SAGA-like hosts since the model updates in the low-mass galaxy regime do not affect the galaxy--halo connection at the MW-mass scale (Fig.~\ref{fig:smhm}). The $M_K$ values for the simulated SAGA-like hosts are obtained by abundance matching the 6dF luminosity function~\citep{2006MNRAS.369...25J} to halo peak maximum circular velocity in c125-2048.}
    \label{fig:host}
\end{figure}

However, this $M_K$ model derives galaxy luminosities independent of the model stellar masses of these hosts from UM. Observed SAGA DR3 hosts have a strong negative correlation (Pearson $r=-0.88$) between $M_K$ and $\mstar$. If we directly use the stellar masses predicted by UM, the correlation with their abundance-matched $M_K$ is only $-0.56$, causing the mock SAGA-host sample to have longer distribution tails than observed SAGA hosts at the smallest and largest stellar masses. To ensure the appropriate correlation between host $M_K$ and $\mstar$, we use 2D Gaussian Kernel Density Estimation~\footnote{\url{https://docs.scipy.org/doc/scipy/reference/generated/scipy.stats.gaussian_kde.html}} (KDE) to interpolate the probability density distribution of observed SAGA hosts in the $M_K$-$\mstar$ plane. We use this KDE as the sampling weight to sub-sample the mock SAGA-like hosts in c125-2048, mimicking the intrinsic $M_K$-$\mstar$ correlation in SAGA. Since we re-sample the mock SAGA hosts without replacement, 2500 out of 7473 is roughly the sub-sample size that populates the observed SAGA host $\mstar$ range well. A sub-sample larger than 2500 will start to include a significant number of mock hosts from the low and high-mass end tails that we wish to exclude.

In Fig.~\ref{fig:host}, we show the final 2500 SAGA-like host sample's stellar mass ($\mstar$) versus halo mass ($M_{\rm vir}$) and $M_K$ distributions in comparison to the observed SAGA hosts. We also show the most recent observationally inferred stellar and halo masses of Milky-Way~\citep{2020MNRAS.494.4291C} for reference (converting their $M_{200}$ to our $M_{\rm vir}$ definition). Since the UM model update mainly affects low-mass galaxies with $\mstar\lesssim 10^9 \msun$, the mean stellar mass of these SAGA-like hosts at a certain halo mass do not change significantly with the UM model update (see Fig.~\ref{fig:smhm}). We have checked that the individual host's stellar masses are within $\sim 0.2$ dex between UM-SAGA and UM DR1, which includes galaxy--halo sampling randomness introduced by re-running the model. Hence, we quote the stellar masses from  UM-SAGA introduced in Section~\ref{sec:results}. We keep this 2500-host catalog \emph{fixed} when selecting SAGA-like satellites (Section~\ref{sec:model:dr1-free}).

Note that the host selection criteria used in this work are almost identical to the host selection in the subhalo abundance matching model~\citep[AM-N20 for short]{2020ApJ...893...48N} used in SAGA DR2~\citep{Mao2021}, with the addition of the brightest neighbor cut and host halo mass cut in our selection. Given that UM DR1 already produces a reasonable SMHM relation that matches the SMHM relation inferred from recent MW satellite observations using an extension of the N20 model~\citep{2020ApJ...893...48N}, using a $v_{\rm peak}$-$M_K$ abundance matching host selection also conveniently enables detailed future comparisons between the UM-SAGA model and AM-N20, which could potentially yield valuable physical insights on low-mass galaxy formation.

\subsection{SAGA satellite selection}
\label{sec:analysis:sats}

Based on the 2500 SAGA-like hosts selected in the previous section, we select subhalos that might host SAGA-like satellites around these mock SAGA-like hosts (Fig.~\ref{fig:model} panel 5). We project all objects in c125-2048 along the $z$ axis of the simulation box. We select SAGA-like subhalos that have a projected distance $10\leqslant R_{\rm host}/\mathrm{kpc}\leqslant 300$  and line-of-sight velocity difference $|\delta v_{\rm LOS}|\leqslant 275\ \mathrm{km\,s^{-1}}$ relative to a nearby SAGA-like host. In the rare occurrence of a subhalo being projected into 300 kpc of two SAGA-like hosts (unlikely due to our isolation criteria), we assign the host halo with the smaller $R_{\rm host}$ as the host of the SAGA-like subhalo. We also impose a halo mass threshold as a resolution cut of $v_{\rm peak} > 35\ \mathrm{km\,s^{-1}}$ for SAGA-like subhalos (see Appendix B in \citealt{2021ApJ...915..116W}). This results in a catalog of 45204 SAGA-like subhalos at $z=0$ that also self-consistently includes interlopers along the line of sight. During model evaluation, we normalize the SAGA-like subhalo mass function by $101/2500$ to compare mock SAGA samples generated in c125-2048 with the corrected SAGA3 satellite SMF around 101 SAGA hosts. 

The subhalo selection described above only includes mock SAGA satellite galaxies occupying subhalos that survive until $z=0$. However, there could be a significant fraction of satellite galaxies whose subhalos already disrupted, i.e., orphan galaxies~\citep{2005Natur.435..629S,2017MNRAS.469..749P,2019ApJ...873...34N,2019MNRAS.488.3143B,2020ApJ...893...48N}. As we discussed in Section~\ref{sec:model:orphan}, orphan galaxies are mostly quenched because their subhalos are heavily stripped. Therefore, the duration for tracing subhalos after their disruption is a crucial parameter that affects the predicted SAGA satellite quenched fractions and needs to be constrained during parameter fitting. Hence, in addition to the $z=0$ surviving SAGA-like subhalo population as mentioned above, we also include disrupted subhalos that potentially host orphan SAGA satellites in our mock sample. Since changing the orphan tracking threshold $T_{\rm merge, 300}$ will change the number and duration of disrupted subhalos that are tracked, the final orphan catalog at $z=0$ will change with every parameter draw of $T_{\rm merge, 300}$. We dynamically select orphan SAGA satellite \emph{galaxies} according to their predicted $z=0$ positions and velocities that satisfy $10\leqslant R_{\rm host}/\mathrm{kpc}\leqslant 300$ and $|\delta v_{\rm LOS}|\leqslant 275\ \mathrm{km\,s^{-1}}$ with respect to the 2500 selected mock SAGA hosts. 

\subsection{SDSS isolated field galaxy selection}
\label{sec:analysis:sdss}

For the SDSS isolated field galaxies, we follow the isolation criteria of \citet{geha2012}, such that every isolated galaxy should be the most massive object within 1.5 Mpc of its vicinity. To select isolated halos from the simulation that might host these isolated galaxies, we construct a K-D tree\footnote{\url{https://docs.scipy.org/doc/scipy/reference/generated/scipy.spatial.cKDTree.html}} for all (sub)halos in c125-2048 for efficient distance querying. We query within a 1.5 Mpc-radius sphere for every halo in the box and select central halos with the largest $M_{\rm vir}$ within their 1.5 Mpc vicinity as `Geha12-like' isolated field halos. We also make a more conservative isolation selection than Geha12 that excludes field halos within $3 R_{\rm vir}$ of a more massive halo (typically group or cluster scale halos with $M_{\rm vir} > 10^{13}\msun$ and $R_{\rm vir} > 500$ kpc) to avoid having splashback objects in the isolated sample. Since the Geha12 sample reaches a redshift of 0.055, we include isolated field halos selected from three simulation snapshots that cover this redshift range in c125-2048: 245429 halos at $z=0$, 245771 halos at $z=0.026$, and 246035 halos at $z=0.054$.

\subsection{Prior choices}
\label{sec:analysis:priors}

For the nine UM DR1 parameters explored in this work, we use their posteriors from UM DR1 as model priors for UM-SAGA. Even though these parameters were not tightly constrained, inheriting the posterior knowledge from UM DR1 still helps UM-SAGA to narrow down the range of meaningful initial guesses, which are not wildly inconsistent with existing higher-mass galaxy constraints.

\begin{figure*}
\begin{center}
    \includegraphics[width=2\columnwidth]{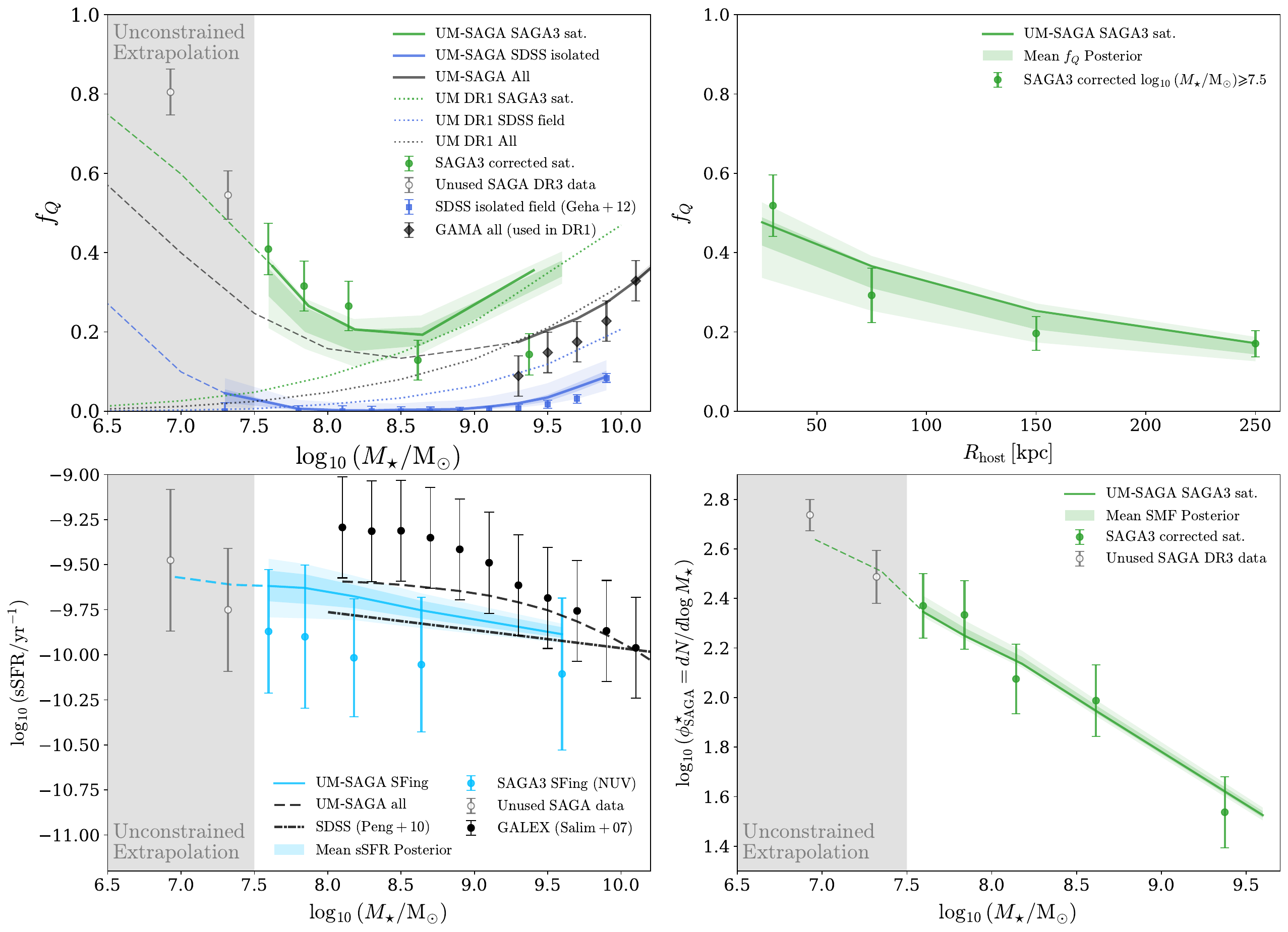}
    \caption{Best-fit UM-SAGA model compared to observational constraints at $z\sim 0$. All SAGA results in this panel have been converted to Chabrier IMF. Solid curves in each panel are the predictions from the best-fit UM-SAGA model in the same mass or radial bins as the corresponding observational constraints. The shaded regions show the $68\%$ (dark) and $95\%$ model posteriors of the mean observables sampling 800 points from the posteriors. The vertical gray-shaded regions represent model extrapolations to $\mstar=10^{6.5}\msun$. {\it Top left}: Quenched fractions versus galaxy stellar mass, $f_Q(\mstar$), for SAGA-like satellites (green), SDSS-like isolated field galaxies (dark blue), and all galaxies (black). Dashed curves of the same color are model extrapolations of $f_Q$ for each galaxy sample outside the mass range where they have observational constraints. The dotted curves are UM DR1 predictions of the three galaxy samples, which significantly underestimate $f_Q$ at $\mstar\lesssim 10^{8.5}\msun$. The left-most black data point from \citet{2013MNRAS.434..209B} represents the minimum mass to which UM DR1 was constrained in $f_Q$. The shaded posteriors around the black curve are negligible, highlighting the non-changing high mass $f_Q(v_{\rm Mpeak})$ part inherited from UM DR1. {\it Top right}: Quenched fraction as a function of 2D projected distance to host. {\it Bottom left}: Average specific SFR as a function of stellar mass. Light blue points are NUV sSFRs of SAGA star-forming satellites (SFing for short), and the light blue solid curve is the UM-SAGA prediction (posterior in blue shaded). The black points are GALEX NUV sSFR~\citep{2007ApJS..173..267S} for all environments, and the black dashed curve is the corresponding UM-SAGA prediction. The green dotted dashed line is the extrapolated linear fit to SDSS H$\alpha$ SFRs in various environments \citep{2010ApJ...721..193P}. {\it Bottom right}: SAGA3 satellite stellar mass function $\phi^{\star}_{\rm SAGA}$ combining satellites from all 101 hosts. }
    \label{fig:fqm}
\end{center}
\end{figure*}

We choose the priors for the six parameters in the new low-mass quenching model as follows. 
Since the high-mass part of $f_Q(v_{\rm Mpeak}, z=0)$ in UM DR1 reaches $50\%$ at $\log_{10} (v_{\rm Mpeak}/\mathrm{km\,s^{-1}}) = 2.23$, and the SAGA quenched fractions have $f_Q\lesssim 20\%$ at $\mstar \sim 10^9\msun$ (Fig.~\ref{fig:fqm}), we impose a prior on $\log_{10} (V_{Q2}/\mathrm{km\,s^{-1}}) < 2.23$ such that the overall quenched fraction can stay below $50\%$ in the transition region of the high-mass and low-mass quenching models. Also, as it is unlikely for any galaxies to form stars in halos with $v_{\rm Mpeak} < 1\ \mathrm{km\,s^{-1}}$ ($M_{\rm peak}\lesssim 10^4 \msun$)~\citep{2020MNRAS.493.1614F,2020ApJ...893...48N,2022MNRAS.516.3944M,2024MNRAS.529.3387A}, we also require $\log_{10} (V_{Q2}/\mathrm{km\,s^{-1}}) > 0$. We do not have prior knowledge on $\sigma_{VQ2, 0}$ apart from the fact that the new low-mass quenching model should have an increasing $f_Q$ with decreasing halo/stellar mass. We thus adopt a wide flat prior for $\sigma_{VQ2,0}\in [0.01, 2]$, where the 0.01 threshold is set to avoid a steep step-like functional form for $f_Q(v_{\rm Mpeak})$. 

Although we do not include higher redshift quenched fraction data in this work, $f_Q(v_{\rm Mpeak})$ is parameterized redshift-dependently to allow for future modeling of higher redshift data, including LG dwarfs SFHs~\citep{2014ApJ...789..147W} and JWST observations. Given the $f_Q$ evolution being monotonically decreasing with increasing redshift for LG dwarfs with $\mstar \lesssim 10^9\msun$\citep{2014ApJ...789..147W}, we impose a prior on the $f_Q = 50\%$ $v_{\rm Mpeak}$ scale to be monotonically decreasing with redshift, i.e., $V_{Q2, a} < 0$ and $V_{Q2, z} < 0$. For the scatter evolution $\sigma_{VQ2,a}$ and $\sigma_{VQ2, z}$ of the new low-mass part of $f_Q(v_{\rm Mpeak})$, we simply assume a wide flat prior between $[-3, +3]$. Although we do not expect these four redshift evolution parameters to be well constrained by the new $z\sim 0$ low-mass galaxy constraints from SAGA and SDSS, we nonetheless leave these parameters free during MCMC to marginalize over all possibilities allowed by the $z\sim 0$ observational constraints. Appendix~\ref{sec:App_A} provides a more detailed summary of prior choices (Table~\ref{tab:A}).

\subsection{MCMC setup}
\label{sec:analysis:mcmc}

The UM-SAGA model generates predictions of SFR($z$) and $\mstar(z)$ for the combination of surviving SAGA-like subhalos at $z=0$, dynamically selected orphan SAGA-like satellites, and SDSS-like isolated field halos. These predictions are then compared to the new low-mass galaxy data described in Section~\ref{sec:data:saga} and \ref{sec:data:sdss}. 
The model also generates predictions for existing UM DR1 constraints (Section~\ref{sec:data:um-dr1}) using the full simulation box (Section~\ref{sec:analysis:sim}). 

To compare with the observed quenched fractions, we define galaxies with model-predicted $\mathrm{sSFR}<10^{-11}\,yr^{-1}$ as quenched and otherwise as star-forming. This definition is largely consistent with the quenching definition of SAGA satellites (\paperfour{}) and SDSS isolated field galaxies~\citep{geha2012}. We convert the SAGA stellar masses and star formation rates from assuming a Kroupa IMF~\citep{2001MNRAS.322..231K} to a Chabrier IMF~\citep{2003PASP..115..763C} by dividing by 1.07 ($-0.029$ dex) when using them as data constraints. This aligns the data constraints with the Chabrier IMF assumption in the stellar mass loss model of UM DR1 that accounts for massive star death over time. Also, existing specific SFR constraints in UM DR1 cover all galaxies regardless of being quenched or star-forming~\citep[e.g., GAMA,][]{2013MNRAS.434..209B}. We explicitly pick out those satellites of SAGA-like hosts that are predicted as star-forming in our simulation compared to the NUV sSFR constraints from star-forming SAGA satellites.

We compute a combined $\chi^2$ for the collection of 1139 data points, 31 points for the new low-mass galaxy data, and 1108 points (shifted values to correct for the change of simulation box, Section~\ref{sec:analysis:sim}) from the existing UM DR1 data, as we vary the 15 UM-SAGA parameters (Section~\ref{sec:model}) while fixing the remaining 37 irrelevant parameters in UM DR1. This is where the selected mock (sub)halo catalogs described in Sections~\ref{sec:analysis:sats}
and \ref{sec:analysis:sdss} are forward modeled into galaxy observable space (SFR and $\mstar$, Section~\ref{sec:model}) and collectively tested against data constraints. We evaluate model precision using the maximum likelihood analysis, setting the likelihood as $\mathcal{L}\sim\exp(-\chi^2/2)$. In most cases, we assume that the observational errors are uncorrelated between different stellar mass or distance bins (e.g., the new stellar mass function and quenched fractions for SAGA satellites); however, we use covariance matrices to evaluate $\chi^2$ wherever available (e.g., the projected two-point correlation functions). The parameter space posterior is efficiently explored using MCMC with a hybrid adaptive Metropolis approach and a stretch-step sampler \citealt{haario2001,goodman2010ensemble}). 

We implement 400 independent walkers, each with 200 burn-in steps, and map out the posterior distribution with 1075 steps after burn-in. For the nine low-mass galaxy-related parameters in UM DR1 that we explore  (Section~\ref{sec:model:dr1-free}), we sample 400 points from their DR1 posteriors as initial positions of their MCMC chains. This effectively takes into account the constrained posterior information on these parameters, given observations already used in UM DR1. As for the six parameters in the new low-mass quenching model we introduced, their initial positions are drawn from their priors.

Our MCMC exploration of the posterior space constrained by the new low-mass galaxy data is well-sampled by $4.3\times 10^5$ points. Our chains are well converged with Gelman--Rubin statistics $R\lesssim 1.2$ and achieve $\gtrsim 5000$ effective samples for each parameter. We summarize the $68\%$ posterior distributions of UM-SAGA and UM DR1 (nine parameters) in Table~\ref{tab:1}; we include the best-fit parameters and adopted priors in Table~\ref{tab:A}. We take the MCMC step in the full posterior MCMC chains with the lowest $\chi^2$ value and perform a brute-force minimum-$\chi^2$ search for the `best-fit' UM-SAGA model around that point, assuming that the lowest $\chi^2$ (reduced $\chi^2$ of 0.36) point from the sampled chains is sufficiently close to the true minimum. In Appendix~\ref{sec:App_B}, we compare the best-fit UM-SAGA model predictions of all UM DR1 constraints and demonstrate that the new model produces nearly identical predictions for existing higher-mass galaxy data constraints after introducing low-mass galaxy quenching in this work.

\section{Results}
\label{sec:results}

This section presents key results of the `best-fit' UM-SAGA model. UM-SAGA well-reproduces the low-mass galaxy population data that was used to constrain it. It also provides novel predictions for low-mass galaxy properties that provide physical insights into low-mass galaxy star formation and quenching. The key takeaways of the UM-SAGA results are:
\begin{enumerate}
    \item UM-SAGA broadly matches the new low-mass galaxy constraints, especially the large difference in $f_Q$ required for SAGA satellites and SDSS isolated field galaxies while matching the SAGA stellar mass function (Fig.~\ref{fig:fqm}, Section~\ref{sec:results:mass-trend}). 
    \item UM-SAGA matches the more radially concentrated profile of quenched satellites (Fig.~\ref{fig:fqr}, Section~\ref{sec:results:distance-trend}).
    \item Despite differences in the star formation histories of low mass galaxies, the SMHM relation in UM-SAGA is almost identical to that in UM DR1 (Fig.~\ref{fig:smhm}, Section~\ref{sec:results:shmr}).
    \item The new $z\lesssim 0.05$ constraints shift the predicted star formation histories earlier by $2~\mathrm{Gyr}$ at $\mstar=10^{7.5}\msun$, resulting in UM-SAGA star formation histories that are more consistent with Local Group galaxies than UM DR1 (Fig.~\ref{fig:t50}, Section~\ref{sec:results:sfh}).
\end{enumerate}

\subsection{Star formation as a function of stellar mass}
\label{sec:results:mass-trend}

In Fig.~\ref{fig:fqm}, we present the $f_{\mathrm{Q}}(\mstar)$ relations, the stellar mass function of SAGA3 satellites, and average specific SFR for star-forming SAGA satellites predicted by the best-fit UM-SAGA model compared to the new low-mass galaxy data constraints. In general, the UM-SAGA model captures the significant difference in quenched fractions between isolated field galaxies and satellites around MW-mass hosts from $\mstar\sim 10^7\msun$ to $\mstar\sim 10^9\msun$, while reproducing the correct galaxy stellar mass functions and average sSFRs. UM-SAGA matching the average SAGA satellite SMF also leads to a correct predicted distribution of satellites per host that is consistent with observations (Fig. 14, \paperthree{}). Thus, the model assumption of tying galaxy SFR to (sub)halo mass and assembly is sufficient to simultaneously quench the appropriate number of satellites and keep the isolated field population with similar stellar masses almost entirely star-forming. This indicates that the impact of environmental quenching in low-mass galaxies can largely be accounted for by dark matter halo growth and stripping down to stellar masses of $\mstar\gtrsim 10^7\msun$.

Compared to UM DR1, the UM-SAGA model results in a significant statistical improvement when fitting the data (e.g., the reduced $\chi^2$ goes from 0.87 to 0.36 in the new model, see also Appendix~\ref{sec:App_A} for an $F$-test of the model improvement), especially in terms of $f_Q(\mstar)$ at stellar mass $\mstar\lesssim 10^{8.5}\msun$ where the satellite quenched fractions begin to rise significantly. Our physically motivated model update thus allows a significant portion of low-mass satellite galaxies to quench. It resolves an outstanding issue in UM DR1 that did not have the flexibility to capture low-mass galaxy quenching~\citep{2021ApJ...915..116W}. UM-SAGA achieves this while preserving the model agreement at higher stellar masses with existing UM DR1 constraints (Fig.~\ref{fig:B1} and \ref{fig:B2}) for all galaxies' $f_Q(\mstar)$ from GAMA~\citep{2013MNRAS.434..209B}, sSFR$(\mstar)$ from GALEX~\citep{2007ApJS..173..267S}, and stellar mass functions from GAMA~\citep{2012MNRAS.421..621B}. The UM-SAGA global stellar mass functions are also consistent with newer GAMA stellar mass functions~\citep{2022MNRAS.513..439D} down to $\mstar \gtrsim 10^7\msun$. 

Although UM-SAGA replicates nearly all the newly added observational constraints within observational errors, two caveats point to potential model limitations. First, the most apparent mismatch is that the new model predicts a $>2\sigma$ higher $f_Q$ for SAGA satellites at $\mstar\sim 10^{9.5}\msun$, corresponding to the highest $\mstar$
data point in the top left panel of Fig.~\ref{fig:fqm}. Second, the model systematically predicts $\sim 0.3$ dex higher average sSFRs than the SAGA (bottom left panel of Fig.~\ref{fig:fqm}). These mismatches represent the least sacrifice UM-SAGA has to make under the current model assumptions to achieve the optimal fit to the collection of newly added low-mass galaxy constraints (Section~\ref{sec:data}). However, a better understanding of these limitations will provide crucial insights for the physical interpretation of the UM-SAGA results, which we discuss in detail in Section \ref{sec:insights} and \ref{sec:discussion}.

Finally, we have an interesting prediction for the quenching status of isolated field low-mass galaxies if we extrapolate the UM-SAGA model to $\mstar\sim 10^{6.5}\msun$ where the model is unconstrained by observational data (the c125-2048 simulation has enough resolution for this extrapolation, see Fig.~\ref{fig:smhm}). We see from the top panel of Fig.~\ref{fig:fqm} that the $f_Q$ of isolated field galaxies start to rise with decreasing mass at $\mstar\lesssim10^{7.5}\msun$, below which the predicted quenched fraction \emph{difference} between SAGA-like satellites and SDSS-like isolated field galaxies are nearly constant ($\sim 40\%$). This is mainly driven by the rapidly rising $f_{Q}$ of SAGA satellites at $\mstar \lesssim 10^{8.5}\msun$ (also true of MW and M31 satellites, \citealt{2015ApJ...808L..27W} and Fig.~\ref{fig:fqm} top panel) and nearly all star-forming isolated field satellites. Below $\mstar\sim10^7\msun$, the $f_Q$ for satellites approach $\gtrsim80\%$ by which point the finite difference in halo assembly ($\Delta v_{\rm max}$) at fixed halo mass ($v_{\rm Mpeak}$) requires isolated field galaxies to quench at a significant fraction before being accreted onto a MW-mass host (see discussions in Section~\ref{sec:insight:physical} and Appendix~\ref{sec:App_C}). Otherwise, it will be impossible to form the highly-quenched $f_Q\gtrsim 80\%$ satellites at $\mstar \lesssim 10^7\msun$.

\subsection{Star formation as a function of distance to the host}
\label{sec:results:distance-trend}

\begin{figure}
\begin{center}
    \includegraphics[width=\columnwidth]{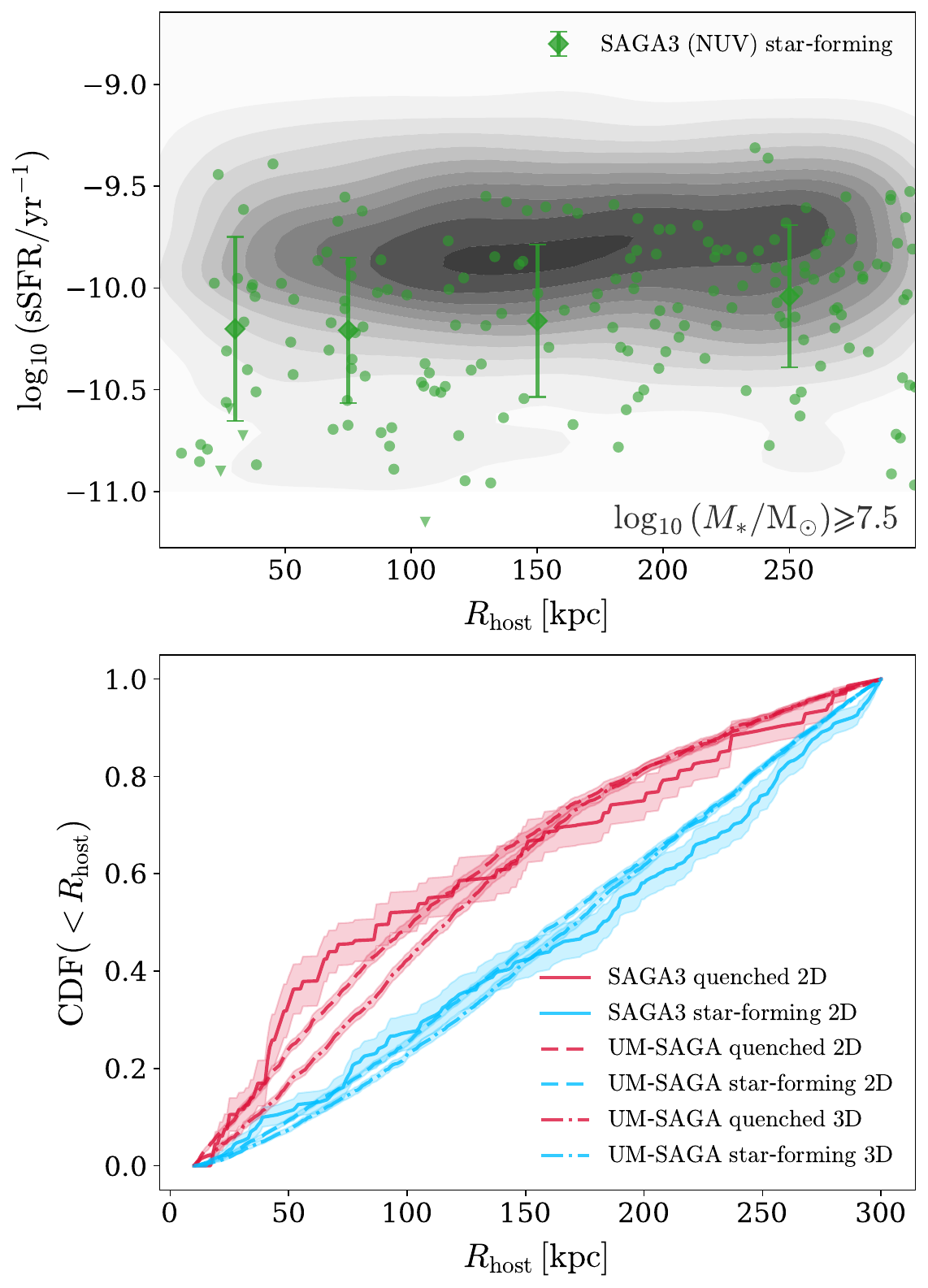}
    \caption{Model predictions of satellite quenching radial trends. SAGA data in this panel were not directly used to constrain the model. {\it Top panel}: Specific SFR versus 2D projected distance to host. Green diamonds with error bars are the average and intrinsic scatter of NUV sSFRs for SAGA star-forming satellites. The scatter points are individual SAGA satellites, with NUV sSFR measurements shown as circles and NUV upper limits shown as triangles. The gray contours show the $10\%$ percentiles of the UM-SAGA prediction. Both the model and the data have rather flat radial sSFR trends. {\it Bottom panel}: Projected 2D cumulative radial number distribution of SAGA quenched (red) and star-forming (blue) satellites. Solid curves are for observations (completeness corrected), and dashed curves are for the model (2D). The dotted-dashed curves show the 3D profiles of true bounded satellites in UM-SAGA. The shaded region around each curve indicates jackknife errors.} 
    \label{fig:fqr} 
\end{center}
\end{figure}

In Fig.~\ref{fig:fqr}, we show the UM-SAGA prediction of satellite star formation quenching as a function of distance to their host.  In the top panel, UM-SAGA captures the mild decrease of satellite-specific SFR towards host centers as in SAGA. In the bottom panel, UM-SAGA predicts a more radially concentrated distribution for quenched satellites than star-forming ones when using both projected and three-dimensional distances, in line with SAGA measurements using projected distance.  Although none of the radial trend observations in this plot were used to constrain UM-SAGA directly, the average sSFR$(\mstar)$ relation indirectly constraints the top panel, while the combination of satellite SMF and $fQ(R_{\rm host})$ indirectly constrains the radial CDFs in the lower panel.

These radial trend predictions help us further understand the model systematics shown in the previous section. The average sSFR over-estimation of $\sim 0.3$ dex by UM-SAGA compared to SAGA is mass-independent and spatially independent. The 0.3 dex higher mean sSFR of UM-SAGA and similar intrinsic scatter to the SAGA star-forming satellites as shown in the top panel of Fig.~\ref{fig:fqr} leads to fewer low SFR star-forming satellites, i.e., `green valley' objects with $\log_{10}(\mathrm{sSFR/yr^{-1}}) \lesssim -10.5$. Furthermore, even though UM-SAGA marginally matches SAGA $f_Q(R_{\rm host})$ within Poisson errors in Fig.~\ref{fig:fqm}, it tends to over-quench satellites in the two middle bins ($50<r/\mathrm{kpc}<200$) while under-quench satellites in the innermost bin ($r< 50$ kpc). This is echoed in the radial CDFs where the quenched UM-SAGA CDF is not as concentrated as SAGA quenched satellites in the inner $\sim 100$ kpc. Since UM-SAGA represents the maximal extent to which quenching radial trends could be explained by dark matter halo assembly, these caveats indicate that additional quenching mechanisms that mainly affect the inner 50 kpc region might be missing UM. Section~\ref{sec:discussion:ssfr} discusses how the host galaxy's lack of a baryonic component may prevent the model from reproducing the steep rise in $f_Q$ and quenched satellite counts as seen in SAGA.

Lastly, the model 3D radial profiles of quenched and star-forming satellites are similar to their 2D profiles in UM-SAGA. The 3D profiles are less concentrated within 150 kpc (by definition) and converge to the 2D profiles beyond 150 kpc. This also indicates that the impact of interlopers projected along the line of sight is small. If interlopers that satisfy the SAGA satellite selection criteria alter the radial profile significantly, removing them should make the 3D CDF more concentrated than the 2D CDF as they dominate large $R_{\rm host}$.   Since interlopers are less stripped than actual bound satellites, they are more likely to survive and avoid becoming an orphan galaxy in UM. As a result, the interloper fraction calculated using the UM-SAGA is $34\%$, marginally higher than the $30\%$ interloper fraction quoted in \paperthree{}, which was based on halo catalogs (\textsc{Rockstar}) without the orphan model applied.

\subsection{Stellar mass--halo mass relation}
\label{sec:results:shmr}

\begin{figure}
\begin{center}
    \includegraphics[width=\columnwidth]{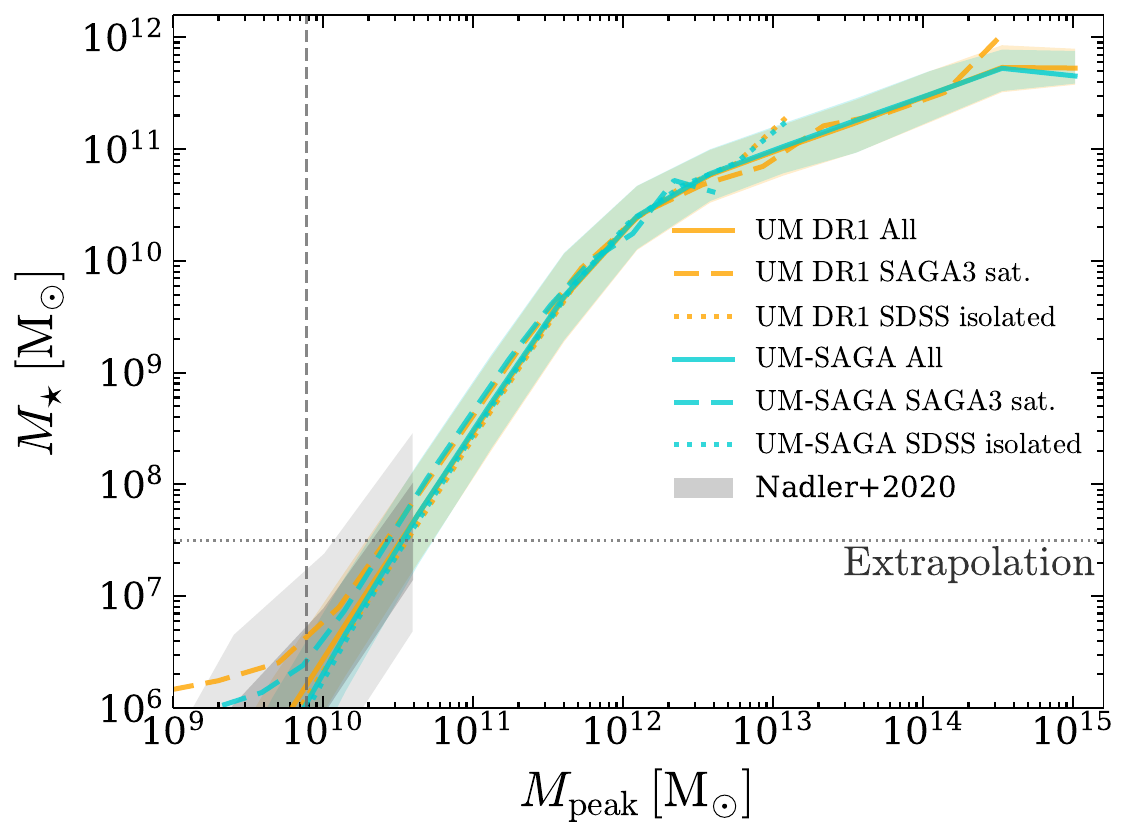}
    \caption{The best-fit UM-SAGA SMHM relation (turquoise) compared to that predicted by UM DR1 (orange) for objects in the c125-2048 cosmological simulation. The solid curves denote all galaxies, the dashed curves denote SAGA-like satellites, and the dotted curves denote SDSS-like isolated field galaxies; shaded bands show $68\%$ scatter in each $M_{\rm peak}$ bin. The UM-SAGA model predicts an almost identical SMHM relation compared to UM DR1 down to $\mstar\sim 10^6\msun$, still consistent with abundance matching constraints inferred from the MW satellite population (gray bands are $68\%$ and $95\%$ marginalized SMHM uncertainties from \citealt{2020ApJ...893...48N}). The horizontal dotted line indicates the stellar mass below which the UM-SAGA SMHM relation is an unconstrained extrapolation. The vertical dashed line is the resolution limit of $300\times$ particle mass in our cosmological simulation.}
    \label{fig:smhm}
\end{center}
\end{figure}

As mentioned in Section~\ref{sec:model:dr1-free}, if the only change made to UM DR1 was the addition of the new low-mass quenching model, the stellar mass halo mass relation could change sufficiently at $\mstar \lesssim 10^9 \msun$ due to earlier quenching in most low-mass galaxies, reducing the average $\mstar$ at fixed halo mass. To self-consistently compensate for this stellar mass suppression and allow the model to match existing SMF constraints at higher masses~\citep{2012MNRAS.421..621B,2022MNRAS.513..439D}, we have explored the redshift-dependent low-mass slope of the mean $\mathrm{SFR}(v_{\rm Mpeak})$ scaling relation in UM to be jointly re-constrained with the low-mass quenching module by the new low-mass galaxy data (Section~\ref{sec:model:dr1-free}).

In Fig.~\ref{fig:smhm}, we show the SMHM relation of the UM-SAGA model (blue) and UM DR1 model (red) for all galaxies (solid), SAGA-like satellites (dashed), and SDSS-like isolated field galaxies (dotted) in c125-2048. The SMHM relation from UM-SAGA is almost identical to that of UM DR1 down to the simulation resolution limit of $M_{\rm peak}\sim 8\times 10^9 \msun$ whether an isolated field galaxy or a satellite. They are also consistent with the SMHM relation inferred using subhalo abundance matching based on recent observations of the MW satellite population (gray bands, AM-N20, \citealt{2020ApJ...893...48N}), despite having different model assumptions, including the treatment of satellite disruption and orphans. This agreement reflects the underlying consistency between the MW and SAGA satellite SMFs (Fig. 8, \paperthree{}) that constrain UM-SAGA and AM-N20. The agreement between UM DR1 and UM-SAGA also indicates that the SAGA and MW SMFs are also consistent with the GAMA SMF~\citep{2013MNRAS.434..209B} down to $\mstar\sim 10^7\msun$ which was part of the UM DR1 constraints~\citep{2019MNRAS.488.3143B}. As we show in Appendix~\ref{sec:App_A} (see Fig.~\ref{fig:A1}) top left corner, the posterior distribution for the 4 $\mathrm{SFR}(v_{\rm Mpeak})$ scaling relation parameters in UM-SAGA are much tighter constrained than the UM DR1 posteriors, which also encompass the UM DR1 best-fit values. This indicates that the change in $\mathrm{SFR}(v_{\rm Mpeak})$ that is required to maintain consistent SMF predictions with higher-mass SMF constraints included in UM DR1 (Fig.~\ref{fig:B1}) is small compared to observational data uncertainties to accommodate for the new degrees of freedom introduced by the new low-mass quenching model added in this work.

\subsection{Star formation histories}
\label{sec:results:sfh}

With the SMHM relation of UM-SAGA being almost identical to UM DR1, as shown above, we present how the star formation histories of low-mass galaxies change quantitatively relative to UM DR1 with the model updates in this work. Since the added low-mass quenching module is only constrained by $z\lesssim 0.05$ SAGA satellites and SDSS isolated field galaxies, the UM-SAGA-predicted SFHs are not fully constrained for higher redshift. Rather, these SFHs are the `most-likely' evolution paths predicted by UM-SAGA that are required to guide the mock galaxies to the $z=0$ $f_Q$ and SMF values consistent with the newly added low-mass galaxy data, marginalized over all possible scenarios allowed by observational uncertainties. We show four example SFHs of a SAGA-like satellites and a SDSS-like isolated galaxies at $\mstar \sim 10^{7.5}\msun$ in Appendix~\ref{sec:App_D}.

Fig.~\ref{fig:t50} shows the distribution of lookback times when the SAGA-like satellite low-mass galaxies in UM-SAGA formed $50\%$ ($t_{50}$) and $90\%$ ($t_{90}$) of their $z=0$ stellar mass. The UM-SAGA model starts to differ from UM DR1 at $\mstar\sim10^{8.5}\msun$, below which $t_{50}$ and $t_{90}$ both systematically shift to earlier cosmic times with decreasing galaxy stellar mass. If we assume $t_{90}$ as a proxy for the galaxy quenching time (e.g., \citealt{2015ApJ...804..136W}), then most of the bright dwarfs at $10^{7.5}\lesssim \mstar/\msun\lesssim 10^9$ quenched within the past 3 Gyrs. We also over-plot the $t_{50}$ and $t_{90}$ times derived from color--magnitude diagrams of individual stars from resolved HST imaging in LG low-mass galaxies~\citep{2014ApJ...789..147W}. The earlier $t_{50}$ and $t_{90}$ values for the UM-SAGA model are qualitatively more favorable than UM DR1 when compared to the LG dwarfs, but further work to quantitatively evaluate the statistical consistency between UM-SAGA and the LG dwarfs SFHs is needed. In the future, we plan to model the reionization quenching of ultra-faint LG dwarfs~(e.g., \citealt{2014ApJ...796...91B,2014ApJ...789..147W}) jointly with the environmental quenching of low-mass galaxies in the SAGA mass range to better understand the transition mass scale of these quenching mechanisms.

\begin{figure}
\begin{center}
    \includegraphics[width=\columnwidth]{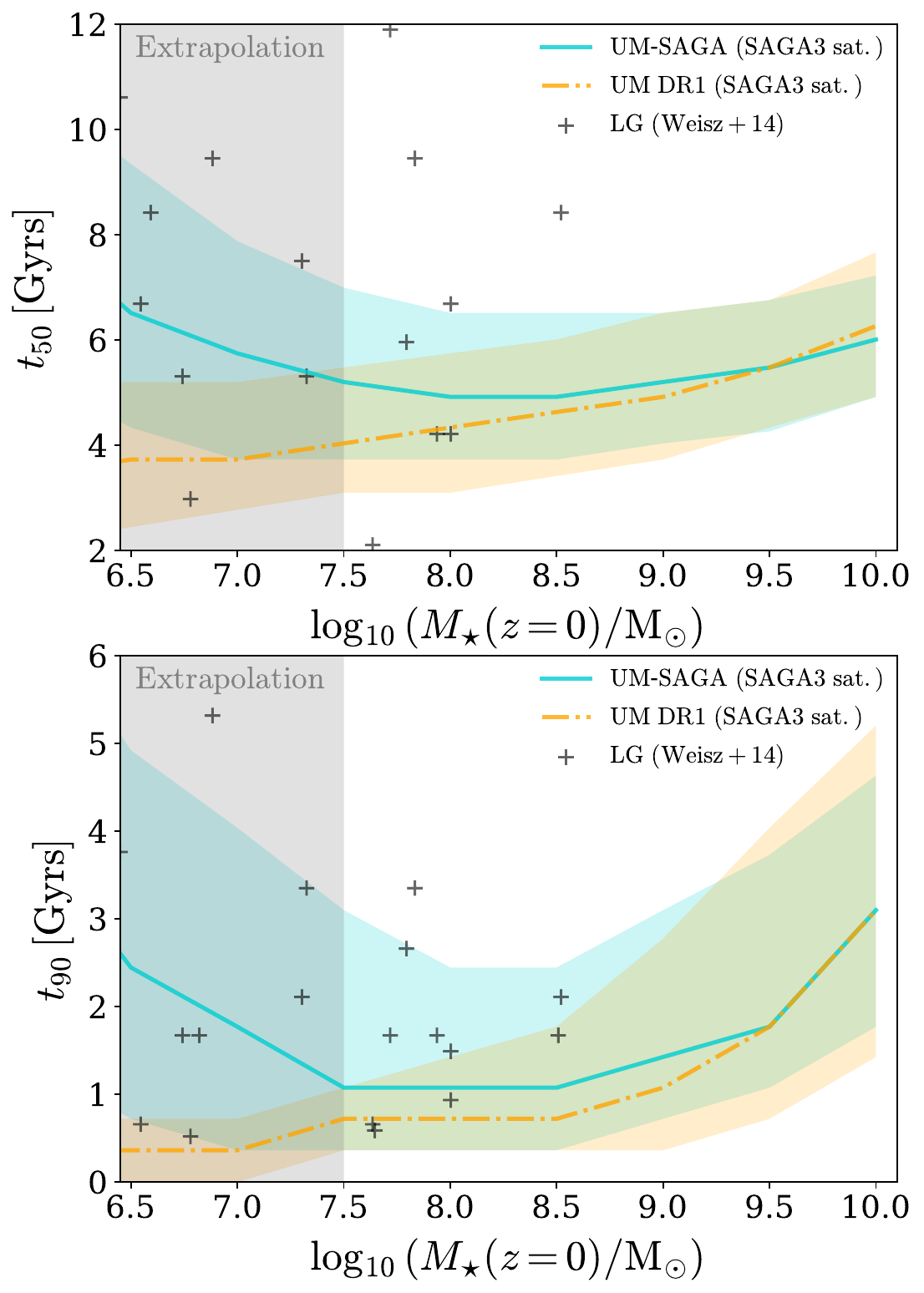}
    \caption{Distribution of lookback times when the SAGA3-like satellites in c125-2048 formed $50\%$ ($t_{50}$) and $90\%$ ($t_{90}$) of their $z=0$ stellar mass in the UM-SAGA model (this work) and UM DR1 model. The solid curve shows the median, while the shaded region shows the $[16\%, 84\%]$ distribution in each stellar mass bin. With the added low-mass quenching model, low-mass satellites systematically form stars and quench earlier in the new model compared to UM DR1 at $\mstar\lesssim 10^8\msun$. This also brings them closer to the observed LG dwarfs ($\gtrsim 70\%$ satellites), which were not used as data constraints in this work (black crosses, \citealt{2014ApJ...789..147W}).}
    \label{fig:t50} 
\end{center}
\end{figure}

\section{Physical insights into environmental quenching}
\label{sec:insights}

This section discusses critical physical insights into environmental quenching gained by constraining UM-SAGA with SAGA satellites and SDSS isolated field galaxies. Their starkly different quenched fractions tightly constrain the model and require the maximum correlation between SFR and halo assembly. Environmental quenching could be solely explained by differences in halo assembly histories down to $\mstar\geqslant 10^{7.5}\msun$. But this drives the model into a corner: if the difference in satellite and field $f_Q$ becomes even larger at lower masses, halo assembly would not be able to explain such large differences and additional quenching mechanisms (e.g., baryonic physics that is not directly connected to this assembly) become necessary. 

\subsection{Impact of SAGA satellites on the relation between SFR and halo assembly}
\label{sec:insights:rc}

As shown in Fig.~\ref{fig:fqm}, the UM-SAGA model can reproduce the large observed difference in quenched fractions for MW-mass host satellites and isolated field galaxies with similar stellar masses between $\mstar\sim 10^{7.5}\msun$ and $\mstar\sim 10^9\msun$. Furthermore, that difference persists when UM-SAGA is extrapolated to $\mstar\sim 10^{6.5}\msun$. Because UM correlates galaxies' SFRs to their halo assembly ($\Delta v_{\rm max}$) rather than relying on satellite-specific processes, and because the SMHM relations for SAGA-like satellites and SDSS-like isolated field galaxies are quite similar (Fig.~\ref{fig:smhm}), the model's success in matching the large difference between satellite and isolated field quenched fractions at similar low-mass galaxy masses is largely due to their (sub)halo differences in assembly history ($\Delta v_{\rm max}$). In this section, we assess how strongly halo assembly and galaxy SFR are correlated (parameterized by $r_c$) as constrained by the new low-mass galaxy data and discuss how future observations of isolated field and satellite dwarfs down to $\mstar\sim 10^{6.5}\msun$ are important for further quantifying the contribution of (sub)halo accretion/stripping to environmental quenching. 

\begin{figure}[t!]
\begin{center}
    \includegraphics[width=\columnwidth]{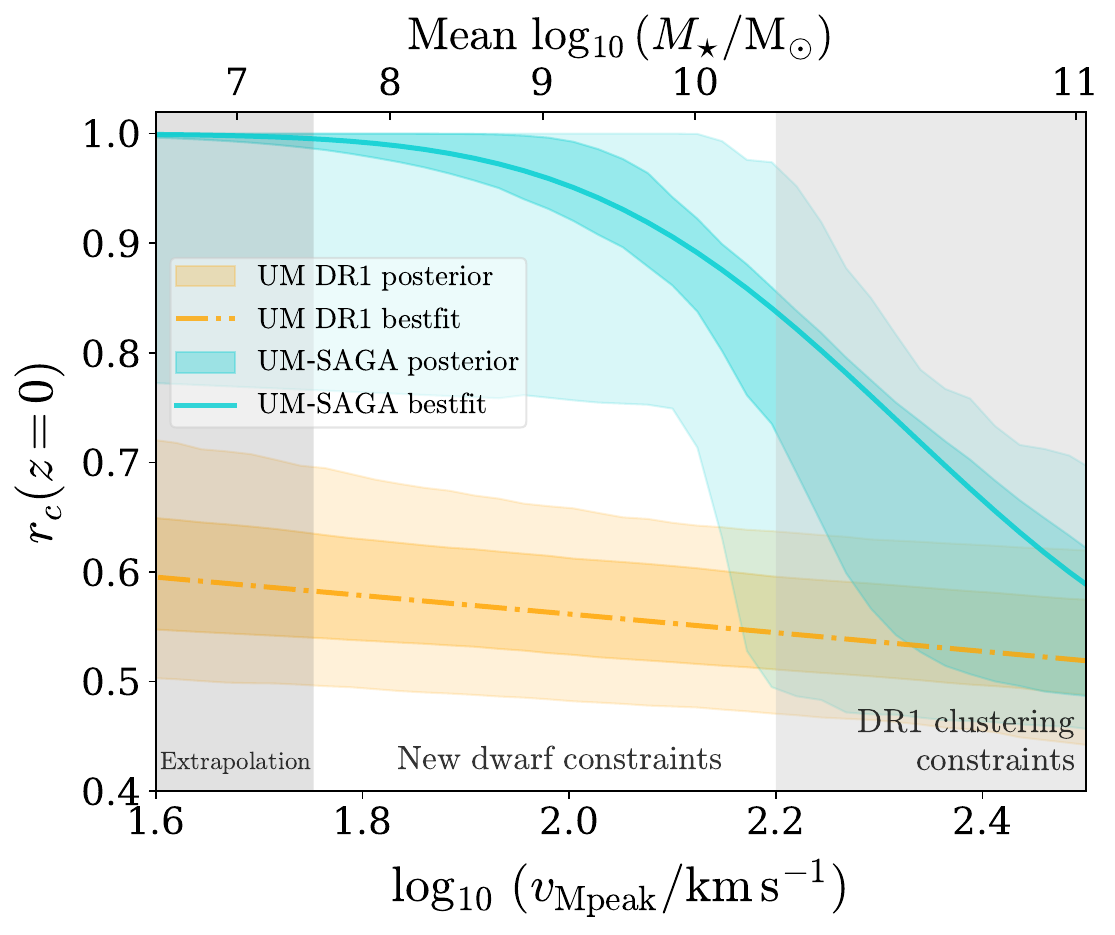}
    \caption{The rank correlation coefficient $r_c(v_{\rm Mpeak})$ between halo accretion ($\Delta v_{\rm max}$) and SFR. Turquoise is the UM-SAGA model, while orange is the UM DR1 (shaded error bars denote darker $68\%$ and lighter $95\%$ distributions of the parameter posteriors). The UM-SAGA $r_c$ approaches the UM DR1 best-fit value in the halo mass range where galaxy clustering constraints were available at $\mstar \geqslant 10^{10.3}\msun$. At the low-mass end, the new model requires $r_c\rightarrow 1$ to simultaneously match the high quenched fractions of SAGA satellites and $f_Q\sim 0$ for SDSS isolated field galaxies. The model is thus approaching the strongest possible correlation between halo assembly and SFR toward low masses.}
    \label{fig:rc_vmp} 
\end{center}
\end{figure}

Fig.~\ref{fig:rc_vmp} shows the best-fit $r_c(v_{\rm Mpeak})$ relation UM-SAGA compared to the best-fit in UM DR1 at $z=0$. At $\log_{10} (v_{\rm Mpeak}/\mathrm{km\,s^{-1}})\gtrsim2.2$ ($\mstar> 10^{10.3}\msun$) where UM DR1 was constrained by two-point correlation functions at $z<0.7$, the $r_c$ in the UM-SAGA model is converged with UM DR1. However, for $\log_{10} (v_{\rm Mpeak}/\mathrm{km\,s^{-1}})\lesssim2.2$ where we have added new low-mass galaxy constraints, the inferred $r_c$ value in UM-SAGA is much higher than that in UM DR1. Even in the transition region of $\log_{10} (v_{\rm Mpeak}/\mathrm{km\,s^{-1}})\sim2.2$ ($\mstar\sim 10^{10}\msun$), $r_c$ is pushed to larger values due to the new data, although the predicted two-point correlation functions change little and only increase slightly at small scales ($r_p<1$ Mpc, Fig.~\ref{fig:B2}). This result indicates that an increasingly strong correlation between galaxy SFR and their (sub)halos' accretion status $\Delta v_{\rm max}$ is required toward lower galaxy masses to simultaneously match the mostly star-forming isolated field galaxies and the significant quenched fraction of satellites around MW-mass hosts. The strong correlation we infer also explains why the best-fit UM DR1 $r_c(v_{\rm Mpeak})$ relation appeared to be linear even though it was parameterized as an error function (Eq.~\ref{eq:7}), as there was no observational data related to the environmental quenching of low-mass galaxies ($\log_{10} (v_{\rm Mpeak}/\mathrm{km\,s^{-1}})\lesssim2.2$) used in the UM DR1 fit, resulting in a poorly-constrained mass dependence of $r_c$.

\subsection{Capturing environmental quenching effects with halo assembly histories}
\label{sec:insight:physical}

\begin{figure*}
\begin{center}
    \includegraphics[width=2\columnwidth]{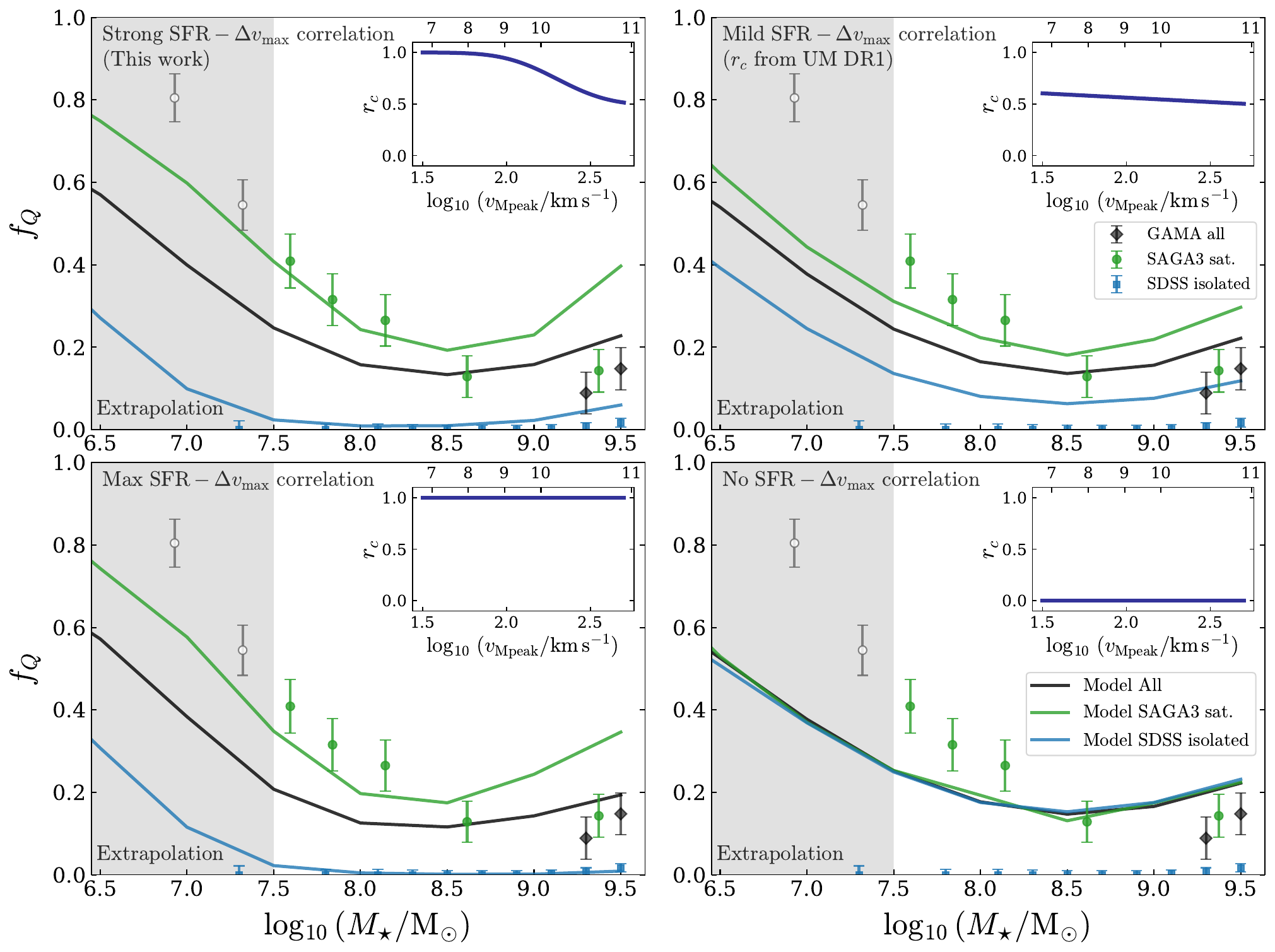}
    \caption{Exploring the effect on satellite and isolated field $f_Q$ due to different correlation strengths between halo growth/stripping and star formation/quenching ($r_c$). In these four panels, we only vary the four $r_c$ parameters (insets) defined in Eq.~\ref{eq:7} and keep the other 11 parameters we explored (including the six in the $f_Q(v_{\rm Mpeak})$ relation, Eq.~\ref{eq:4}) fixed to their best-fit values in this work. The top axis for the inset panels is the corresponding stellar mass at different $v_{\rm Mpeak}$.  {\it Top left:} The best-fit $r_c$ in this work (blue curve in Fig.~\ref{fig:rc_vmp}). $r_c \rightarrow 1$ at $v_{\rm Mpeak} \lesssim 100\,\rm{km\,s^{-1}}$, $\mstar \lesssim 10^9 \msun$. {\it Top right:} Using the $r_c$ values from UM DR1, which is the green curve in Fig.~\ref{fig:rc_vmp}. With an $r_c\sim 0.5$, the SAGA-satellite $f_Q$ differs from the isolated field $f_Q$, but their differences are too small compared to observations. {Bottom left:} Setting $r_c = 1$, maximally allowed correlation between halo stripping and galaxy quenching, similar to UM-SAGA. {\it Bottom right:} Setting $r_c = 0$, no correlation between halo stripping and galaxy quenching, SFR only depends on halo mass. The model $f_Q$ for isolated field galaxies is identical to SAGA-like satellites in this case, both being the same as $f_{Q, \rm all}$. Comparing the two left panels, UM-SAGA almost requires the maximally allowed correlation strength between SFR and $\Delta v_{\rm max}$ to capture the large difference in $f_Q$ for SAGA satellites and isolated field galaxies.}
    \label{fig:rc} 
\end{center}
\end{figure*}

To further illustrate the effect of $r_c$, Fig.~\ref{fig:rc} shows the galaxy quenched fractions under four different choices of $r_c(v_{\rm Mpeak})$ parameters (Eq.~\ref{eq:7}), keeping all other parameters fixed. The goal of this experiment is to demonstrate how different SFR--halo assembly correlations manifest as environmental quenching (differences in satellites/isolated $f_Q$), and to reveal the limitations of modeling SFR for LMC-mass satellites through $\Delta v_{\rm max}$. The smaller the absolute value of $r_c$, the larger the scatter between SFR and halo assembly becomes at fixed halo mass, according to Equation~\ref{eq:8}. We consider the following four cases in Fig.~\ref{fig:rc}:
\begin{enumerate}
    \item \emph{UM-SAGA best-fit model}: The top left panel shows the best-fit parameters in the UM-SAGA model, with the $f_Q$ curves being identical to Fig.~\ref{fig:fqm} top panel and the inset $r_c$ curve being identical to the blue curve in Fig.~\ref{fig:rc_vmp}. The $r_c\rightarrow 1$ at $\log (v_{\rm Mpeak}/\mathrm{km\,s^{-1}})\lesssim 2$ allows for a larger separation in satellite versus isolated field $f_Q$ at $\mstar\lesssim 10^{8.5}\msun$, in agreement with the new data from SAGA and SDSS.
    \item \emph{Maximal correlation between halo assembly and SFR}: The bottom left panel sets $r_c = 1$ to demonstrate the theoretical maximum separation of $f_Q$ between SAGA-like satellites and SDSS-like isolated galaxies under the current UM framework, which is quite similar to the best-fit UM-SAGA model. The slight differences in $f_Q$ in the bottom right compared to the bottom left result from removing scatter in the SFR--$\Delta v_{\rm max}$ ranking ($R_N$ in Eq.~\ref{eq:8}) when $r_c$ is set strictly to 1, slightly changing the stellar masses of the model galaxies.
    \item \emph{UM DR1 halo assembly--SFR correlation}: The top right panel uses the best-fit $r_c$ parameters from UM DR1 (red curve in Fig.~\ref{fig:rc_vmp}). The $r_c\sim 0.5$ value (constrained by two-point correlations at $\mstar >16^{10.3}\msun$ from SDSS in UM DR1) in this scenario predicts a difference in $f_Q$ for SAGA-like satellites and SDSS-like isolated field galaxies, but their differences are too small compared to the new low-mass galaxy constraints, especially at $\mstar\leqslant 10^{8.5}\msun$.
    \item \emph{No correlation between halo assembly and SFR}: The bottom right panel sets $r_c = 0$ across all redshifts, eliminating any correlation between galaxy SFR and halo assembly and only keeping the halo mass dependence of galaxy SFR. This erases all environmental dependence of galaxy quenching and produces identical $f_Q(\mstar)$ relations for SAGA-like satellites and SDSS-like isolated field galaxies, which both converge to the model $f_{Q, \rm all}$ in this case ($\mstar > 10^9\mstar$). This is the \emph{only} scenario in the current UM framework that can simultaneously correctly produce the $f_Q$ for LMC-mass galaxies in the overall galaxy population and around SAGA-like hosts. However, this model is ruled out by higher mass clustering constraints and lower mass quenched fraction constraints from SAGA, thus demonstrating a potential model inflexibility (see Section~\ref{sec:discussion:lmc} for more discussion).
\end{enumerate}

Fig.~\ref{fig:rc} emphasizes that nearly the maximum allowed correlation between halo stripping and galaxy quenching ($r_c\rightarrow 1$) is required to match the new low-mass galaxy constraints added in this work. Specifically, SAGA provides crucial constraints for satellites around MW-mass hosts that drive the \emph{overall} quench fraction up at $\mstar\lesssim 10^{8.5}\msun$. This results in the $f_Q \gtrsim 30\%$ prediction for isolated field galaxies when UM-SAGA is extrapolated down to $\mstar = 10^{6.5}\msun$, which represents the lower bound requirement on isolated field $f_Q$ that needs to be achieved under the model assumption that halo assembly fully determines galaxy SFR. 

\subsection{Physical interpretation of a strong SFR-halo assembly correlation}
\label{sec:insight:physical}

The physical interpretation of our $r_c\longrightarrow 1$ result is anchored to the assumptions that make up UM, i.e., galaxy formation and evolution at different masses and environments are primarily driven by variations in halo mass (primary bias $v_{\rm Mpeak}$) and halo assembly (secondary bias $\Delta v_{\rm max}$). How well the model reproduces the input observational constraints reflects how flexible the model is under this set of physically motivated assumptions. The extreme scenario of galaxy SFR having to drop immediately as it becomes a subhalo and starts to tidally strip a strong condition for quenching timescales (Fig.~\ref{fig:t50}). Even though it seems extreme, it is qualitatively consistent with hydrodynamic simulation predictions of rapid loss of a low-mass galaxy’s gas reservoir following its accretion into a MW-mass host \citep[e.g., Fig. 11] []{2023MNRAS.522.5946E}. Recent hydrodynamic zoom-in simulations~\citep{2024ApJ...961..236C} with sufficient resolution in this regime showed that stellar feedback could cause isolated field galaxies to quench at $\gtrsim 20\%$ while also predicting a $\sim 40\%$ higher $f_Q$ in satellites at $\mstar\lesssim 10^7\msun$ due to environmental quenching, in line with the UM-SAGA predictions.

This $r_c\longrightarrow1$ scenario is needed to reproduce the new low-mass galaxy data under the specific set of UM assumptions, which should be interpreted as data-driven and not necessarily physical. As shown in Fig.~\ref{fig:rc}, it is impossible to explain the SAGA satellite and SDSS isolated $f_Q$ with an $r_c$ that is significantly less than one on the low-mass end. The physical insight behind this is that the difference in star formation properties of the low-mass galaxies are so stark in satellites versus isolated ones that it is almost impossible to explain their differences based on the diversity of halo assembly $\Delta v_{\rm max}$  alone, given their similar halo masses. We show in Appendix~\ref{sec:App_C} that there is only a limited difference in $\Delta v_{\rm max}$ distributions for halos hosting isolated and satellite galaxies with $\mstar \lesssim 10^{8.5}\msun$, which places an upper limit on the difference in $f_Q$ of $\sim 40\%$ for isolated and satellite low-mass galaxies. In this case, if isolated field galaxies at $\mstar\lesssim 10^7\msun$ do not achieve $f_Q\gtrsim 30\%$ before infall, the tidal stripping of their dark matter halos would not be efficient enough to create the $f_{Q, \rm sat}\gtrsim 80\%$ highly-quenched satellite population around MW-mass hosts both seen in SAGA and the Milky-Way itself at this mass scale.

In summary, UM-SAGA demonstrates the maximum extent to which the environmental dependence of low-mass galaxy quenching can be attributed to dark matter halo mass and assembly. The SAGA and SDSS observations are very constraining, and in the context of the current UM framework, this pushes $r_c$ to 1. This result does not rule out the contribution to satellite quenching from other mechanisms such as stellar feedback and ram pressure stripping~\citep{1972ApJ...176....1G,2003AJ....125.1926G,2009ApJ...696..385G,2023MNRAS.522.5946E}. Integrating these baryonic effects into UM and exploring halo assembly parameterizations beyond $\Delta v_{\rm max}$ in the future are important for better understanding the physical context and limitations of UM-SAGA (see Section~\ref{sec:discussion:baryons} for a discussion).

\subsection{Future tests of UM-SAGA predictions}
\label{sec:insight:future}

If future observations such as DESI BGS~\citep{2023AJ....165..253H} and LOW-Z~\citep{2023ApJ...954..149D} indicate that the majority of $\mstar\sim 10^6\msun$ isolated field galaxies quench at $f_Q\ll 30\%$ (e.g., $\sim 10\%$), then satellite quenching must also include contributions from other quenching mechanisms beyond gravity-only sources linked to halo assembly as assumed in UM-SAGA. As mentioned above, additional baryonic quenching mechanisms such as ram-pressure stripping could fill the quenching gap between isolated field and satellite galaxies by efficiently peeling off the gas reservoir of satellites in MW-mass hosts after infall, as recently shown in various hydrodynamic simulations~\citep{2022MNRAS.514.5276S,2023MNRAS.522.5946E}.

So far, the largest known sample of `field' low-mass galaxies at $\mstar<10^7\msun$ is the small number of dwarfs just outside the virial radii of either the MW or M31 in the Local Group~\citep{2014ApJ...792..141S}. This sample, which is less isolated than low-mass galaxies in \citet{geha2012}, allows for a maximum $f_Q\lesssim 40\%$ at $\mstar\sim 10^{6.5}\msun$ while taking into account observational incompleteness and small number statistics, which is consistent with the UM-SAGA prediction (despite slightly different isolation criteria). There are a handful of known {\em isolated} field low-mass galaxies at $\mstar<10^7\msun$  that are quenched and share similar isolation criteria as \citet{geha2012}, i.e., KKR25~\citep{2012MNRAS.425..709M}, Tucana B~\citep{2022ApJ...935L..17S}, COSMOS-dw1~\citep{2021ApJ...914L..23P}, and PEARLSDG~\citep{2024ApJ...961L..37C}, providing further support for the presence of quenched isolated low-mass galaxies. A dedicated observational campaign focusing on the star formation status of isolated field galaxies at $\mstar/\msun \in [10^6, 10^7]$ is strongly motivated to better understand the underlying physics of low-mass galaxy quenching and to better quantify the transition between the late-time environmental quenching in classical dwarfs and the presumed dominance of reionization quenching in ultra-faint-dwarfs at $\mstar\lesssim 10^5\msun$ (see e.g., Fig.~5 of \citealt{2019MNRAS.483.4031R}).

Another important verification for the UM-SAGA predictions is on observed field galaxies with distances to their hosts that are between SAGA satellites and SDSS isolated galaxies, i.e., field galaxies that are roughly $300\sim900$ kpc away from SAGA hosts. In future work, we plan to compare UM-SAGA predictions with the SAGA-bg~\citep{2024ApJ...966..129K} sample, a background galaxy spectroscopic sample generated in the process of identifying SAGA satellites. This sample obviously shares similar completeness characteristics as SAGA satellites and goes down to brighter limiting $\mstar$ compared to the Local Volume field galaxies mentioned above. However, this sample will still provide a valuable test of UM-SAGA, especially in terms of how well halo assembly (parameterized by $\Delta v_{\rm max}$) models the transition region at $R_{\rm vir}\sim 3 R_{\rm vir}$ over a diverse set of MW-mass hosts.

\section{Discussion}
\label{sec:discussion}
Here we discuss UM-SAGA's predictions, caveats, and potential future improvements that will further broaden our understanding of satellite quenching. We discuss the over-estimation of average sSFR in Section~\ref{sec:discussion:ssfr}, the radial trends of satellite quenching (Section~\ref{sec:discussion:baryons}), and the over-quenching of LMC mass satellites (Section~\ref{sec:discussion:lmc}).

\subsection{Impact of quenching timescales on present-day specific SFRs}
\label{sec:discussion:ssfr}

Although UM-SAGA captures the anti-correlation of sSFR versus $\mstar$ as observed in the SAGA satellites, it predicts an sSFR distribution almost $1\sigma$ higher than the average observed value for star-forming SAGA satellites, considering the intrinsic scatter (Fig.~\ref{fig:fqm} bottom left panel). This higher mean sSFR leads to UM-SAGA lacking in reproducing the population of `green valley' star-forming satellites in the $\log_{10} (\mathrm{sSFR/yr^{-1}})\in [-11, -10.5]$ range, which is exemplified in the top panel of Fig.~\ref{fig:fqr}.

The UM-SAGA specific SFRs are optimized to simultaneously predict the higher sSFRs~\citep[GALEX][]{2007ApJS..173..267S} for all galaxies living in diverse environments at $\mstar\gtrsim 10^{8.5}\msun$ and the lower sSFRs in star-forming satellites around SAGA hosts (blue points in Fig.~\ref{fig:fqm}). The main differences in sSFR between SAGA and \citet{2007ApJS..173..267S}, which was also shown in \paperfour{} (Fig. 8 right panel, Leroy+2019 adopts \citealt{2007ApJS..173..267S}),  comes from two factors: i) GALEX covers more diverse galaxy environments with many being in the field, more isolated than SAGA; ii) GALEX  covers higher redshift galaxies out to $z\sim 0.1$ and the average sSFR increases with redshift (e.g., Fig.~\ref{fig:B1}). Apart from these major physical differences, GALEX  Survey's NUV SFR may also differ from SAGA's due to the different dust attenuation models~\citep{2000ApJ...539..718C}. Originally in UM DR1, a constant 0.28 dex uncertainty was given to all mean sSFR constraints to reconcile the large inter-literature differences, which is usually much larger than the reported uncertainties on the mean (see Appendix C3 in \citealt{2019MNRAS.488.3143B}). Given the observational systematics between SAGA and \citet{2007ApJS..173..267S} sSFRs, we follow the treatment of sSFR uncertainties in UM DR1 and assume the {\it intrinsic scatter} in SAGA NUV sSFRs as uncertainties on the mean sSFR (Fig.~\ref{fig:fqm} bottom left), which serves as a conservative error estimate.

We notice that the all-galaxy average sSFR prediction by UM-SAGA (orange-dashed curve in Fig.~\ref{fig:fqm}) is $\sim 0.2$ dex higher than its SAGA star-forming satellite prediction, meaning that the model has enough flexibility to learn the environmental and redshift differences of the two observational datasets. However, similar to the $r_c$ variation experiment in Section~\ref{sec:insight:physical}, we fail to create a larger difference in the predicted sSFRs of the two samples, one for all galaxies and another of SAGA satellites, when we vary $r_c$. We have tried including and excluding SAGA star-forming satellite sSFRs as model constraints; we found negligible change for the average sSFR prediction in the final fitted model. This means that under current UM SFR-$\Delta v_{\rm max}$ rank correlations, it is impossible to create a significantly lower average sSFR for SAGA satellites relative to all galaxies given the limited difference in their halo assembly, and most likely, satellite-specific processes are required.

In addition to NUV-specific SFRs, we also compare UM-SAGA to SDSS H$\alpha$ specific SFRs~\citep{2010ApJ...721..193P} for galaxies in all environments in Fig.~\ref{fig:fqm}. H$\alpha$ is sensitive to star formation in the past $5\sim10$ Myrs, and NUV probes the past $\sim 100$ Myrs. UM-SAGA predictions lie between these two probes at $\mstar\gtrsim 10^{8.5}\msun$, below which SAGA data fill a large gap in previous spectroscopic coverage. We have verified that the UM-SAGA sSFR predictions in Fig.~\ref{fig:fqm} do not change significantly if we use the average SFR between the last two snapshots of c125-2048 near $z=0$ which are set apart by 360 Myr, comparable to the NUV SFR timescale. This suggests that UM-SAGA also predicts star formation to be sustained at $\gtrsim 100$ Myrs in star-forming galaxies, in agreement with quenching timescales inferred for SAGA satellites in \paperfour{} by combining H$\alpha$ and NUV-specific SFRs.

Combining this timescale prediction with the fact that UM-SAGA over-predicts average specific SFR for star-forming satellites, we conclude that UM-SAGA keeps satellites star-forming for roughly the right amount of time but shuts their star formation down too abruptly once they become quenched. Currently, the abrupt quenching process without an intermediate phase of decreased sSFR but still being classified as `star-forming' ($\log_{10}(\mathrm{sSFR/yr^{-1}})>-11$) is likely the reason why UM-SAGA does not produce sufficient `green valley' objects as shown in Fig.~\ref{fig:fqr}. This issue is related to the over-quenching issue of LMC-mass satellites (Fig.~\ref{fig:fqm} top left) further questions the flexibility of $\Delta v_{\rm max}$ for capturing the reduced sSFR of star-forming satellites compared to isolated galaxies at similar masses (see Section~\ref{sec:discussion:lmc} more discussion).

\subsection{Interplay between quenching and satellite disruption}
\label{sec:discussion:baryons}

As discussed in Section~\ref{sec:results:distance-trend}, UM-SAGA does not create a steep enough radial trend in the increase of $f_Q$ and quenched satellite number counts within 50 kpc of their hosts as compared to SAGA quenched satellites. There are likely additional quenching mechanisms near the host center that are inadequately modeled in UM-SAGA, whose details may involve a complex interplay between various baryonic physics, e.g., subhalo disruption due to the central galaxy potential, ram-pressure stripping, shock heating, etc. Quantifying the effects due to processes beyond gravity, such as ram-pressure stripping, requires direct simulation or more detailed modeling. However, gravity-only effects such as disk disruption of subhalos could be modeled more faithfully with better-crafted gravity-only simulations. 

We note that the best-fit $T_{\rm merge, 300}$ increased from 0.63 in UM DR1 to 0.70 in UM-SAGA (Table~\ref{tab:A}); this change is significant given the uncertainties on $T_{\rm merge, 300}$ and results in fewer orphan galaxies being traced to match the newly added SAGA and SDSS data. This model change may partially reflect the higher resolution in c125-2048 (UM-SAGA) relative to {\em Bolshoi-Planck} (UM DR1), along with the constraining power of satellite-specific SMF and quenched fractions from SAGA on $T_{\rm merge, 300}$ given that the orphan model primarily affects satellite predictions. As we mentioned in Section~\ref{sec:model:orphan}, a mock galaxy is considered disrupted if its subhalo is stripped below the orphan tracking threshold $T_{\rm merge, 300}$, even if its subhalo is still surviving. This acts as a simple, effective model in UM-SAGA that considers disk disruption of satellite galaxies. Interpreting the increase in $T_{\rm merge, 300}$ from this perspective suggests that SAGA satellite data requires slightly more satellite disruption in MW-mass hosts than higher-mass galaxy constraints in UM DR1. However, we caution that this argument is based on our current gravity-only simulations, in which the absolute value of $\Delta v_{\rm max}$ of subhalos near host halo centers could be biased low relative to real satellite galaxies due to the lack of tidal stripping from a central galaxy potential.

To further elaborate on the radial trends of satellite quenching, future work based on the EDEN-Symphony simulations~\citep{2024arXiv240801487W} will explore zooming-in on MW-mass hosts with embedded analytic disk potentials~\citep{2010ApJ...709.1138D,2017MNRAS.471.1709G,2019MNRAS.487.4409K}. This will address whether subhalo disruption with more realistic tidal stripping can account for the quenched fraction differences at host halo centers between UM-SAGA and SAGA observations. 

\subsection{Comparing satellite and field quenched fractions at the LMC-mass scale}
\label{sec:discussion:lmc}

The outstanding issue of the UM-SAGA model is that it overestimates the quenched fraction of LMC-mass ($\mstar\gtrsim 10^{9}\msun$) SAGA-like satellites at the high-mass end. Since this mass scale ($\mstar\gtrsim 10^9\msun$) was already constrained by GAMA $f_Q$ data in UM DR1, it is not a novel issue introduced in this work. The green dotted curve in the top panel of Fig.~\ref{fig:fqm} at $\mstar\sim 10^{9.5}\msun$ already predicts a high $f_Q$ for SAGA satellites before the model update. 

As discussed in Section~\ref{sec:insight:physical}, a non-zero $r_c\gtrsim 0.5$ is required by two-point correlation functions at $\mstar>10^{10.3}\msun$ (included in UM DR1) and by the fact that SDSS isolated field galaxies~\citep{geha2012} are almost entirely star-forming up to $\mstar\sim 10^{9.5}\msun$. However, LMC-mass satellites around observed SAGA hosts have similar $f_Q$ to the overall quenched fraction of galaxies in the GAMA Survey (see top panel of Fig.~\ref{fig:fqm}). According to Fig.~\ref{fig:rc}, the only scenario under which the current UM framework can produce $f_Q$ of LMC-mass satellites (green) similar to the overall quenched fraction (black) is the $r_c = 0$ case, which is disfavored by other observational constraints that support $r_c\sim 0.5$ at these masses. Therefore, the combined data set of SAGA satellites and SDSS isolated field galaxies calls for incompatible $r_c$ parameters at $\mstar\sim 10^{9.5}\msun$ and highlights inflexibility in the UM framework to faithfully describe LMC-mass galaxy quenching.

To mitigate this issue, one might argue for refitting the high-mass part ($\mstar\gtrsim 10^{9}\msun$) of the UM $f_Q(v_{\rm Mpeak})$ relation, which is frozen in this work, to bring down the overall $f_Q$ at $\mstar\sim 10^{9.5}\msun$ towards the GAMA $f_Q$ data values. If the model $f_{Q, \rm All}$ is lowered to $\sim 10\%$ by hand and becomes more consistent with GAMA, the LMC-mass satellite $f_Q$ decreases as expected. However, the decrease in $f_{Q, \rm All}$ at $\mstar\sim 10^9$ will ultimately worsen the model predictions for galaxy quenching at $\mstar\gtrsim 10^{10}~\msun$ which is well-constrained in UM DR1. On top of that, as galaxy clustering constraints favor a non-zero $r_c\gtrsim 0.5$ at $\mstar\gtrsim 10^{10}\msun$, the model still predicts a $> 1\sigma$ higher $f_Q$ for LMC-mass satellites than SAGA even if $f_{Q, \rm All}$ is brought to $\sim 10\%$ as in GAMA. 

Therefore, we are motivated to explore beyond the current UM framework of $P(\mathrm{SFR} | v_{\rm Mpeak}, \Delta v_{\rm max}, z)$ in the future. We believe it is quite difficult to swap out $\Delta v_{\rm max}$ entirely with these alternative parameterizations, given how well UM-SAGA reproduces the myriad of observational constraints over four decades in $\mstar$. However, these new parameterizations in conjunction with the $\Delta v_{\rm max}$ methodology may provide simple and physically motivated solutions to the UM-SAGA limitations of over-quenching LMC-mass satellites (Fig.~\ref{fig:fqm}) and lacking `green valley' star-forming satellites (Fig.~\ref{fig:fqr}) by increasing the dynamic range of halo assembly timescales. Potential augmentations to the UM model may include:
\begin{itemize}
    \item $\Delta v_{\rm max}$ measures changes in the halo mass within roughly the scale radius. Including a free physical scale parameter (e.g. $\Delta v(r)$) within which halo assembly is measured can potentially enable more accurate modeling of a galaxy's SFR response to halo accretion or stripping.
    \item Exploring parameterizations of halo assembly beyond $\Delta v_{\rm max}$, such as local tidal forces or orbital pericenter distances (which are specific to satellites). In fact, initial tests combining tidal forces with $\Delta v_{\rm max}$ show promising aspects for resolving the LMC-mass satellite over-quenching issue.
    \item Explicitly introducing satellite--host conformity would allow satellites' SFRs to more strongly correlate with their host halo mass or host galaxy SFR.
    \item Introducing modules accounting for baryonic effects such as stellar feedback and ram-pressure stripping would add flexibility to the model, capturing variability in SFRs that is not purely captured by dark matter halo properties.
\end{itemize}

These potential augmentations are crucial for improving the physicality of the UM framework and may eventually need not require $r_c\longrightarrow 1$ to reproduce the observations. Although we leave a full systematic study of these upgrades to the future, one must eventually understand the new systematics brought along with the new degrees of freedom. One may also need to explore better summary statistics beyond galaxy number densities and correlation functions~\citep{2024arXiv240301393K}, as well as carefully calibrating against a broader range of archival data (e.g., \citealt{2012MNRAS.424..232W}) and incoming observations (e.g., JWST; ~\citealt{2023ApJ...956L..42S}) covering more diverse cosmic environments and cosmic times. These new parameterizations could also shed light on UM model limitations found at other galaxy masses and redshifts~\citep[e.g., clustering of DESI $z\sim 1$ Emission Line Galaxies;][]{2023arXiv231009329Y}. If these augmentations for better quantifying halo assembly fail to resolve the UM-SAGA issues, it would strongly suggest satellite-specific baryonic processes (e.g., ram-pressure stripping), rather than just halo assembly and stripping, as drivers of satellite star formation and quenching.

\section{Summary}
\label{sec:summary}

In this paper, we used the empirical galaxy--halo connection model \textsc{UniverseMachine}~(UM, \citealt{2019MNRAS.488.3143B}) to critically assess the relation between halo assembly and the star formation and quenching of satellite and isolated dwarf galaxies. We added a new redshift-dependent low-mass quenching model to UM that allows low-mass galaxies, especially low-mass satellites around MW-mass hosts, to quench efficiently,  thus resolving a major inflexibility in the original UM DR1 model. The new model, UM-SAGA, is the first version of UM constrained by observations down to $\mstar\gtrsim 10^7\msun$, including the average stellar mass functions, quenched fractions, and NUV-specific SFRs of SAGA satellites, as well as the quenched fractions of isolated field galaxies from SDSS~\citep{geha2012}. Given the new observational constraints, we fit the optimal parameters for UM-SAGA to these data by running adaptive MCMC on mock galaxy catalogs generated from a cosmological gravity-only simulation. Our main findings are as follows:
\begin{enumerate}
    \item Halo assembly can capture environmental quenching by simultaneously matching the high quenched fractions of SAGA satellites while keeping the isolated field galaxies star-forming at $\mstar/\msun\in [10^{7.5}, 10^9]$ (Fig.~\ref{fig:fqm}).

    \item UM-SAGA predicts a more radially concentrated distribution for quenched satellites than star-forming ones, in line with SAGA observations. (Fig.~\ref{fig:fqr}).
    
    \item The SMHM relation of UM-SAGA is almost identical to UM DR1, and both are consistent with the SMHM relation inferred from MW satellites using abundance matching  (Fig.~\ref{fig:smhm}).
    
    \item The large difference in $f_Q$ between SAGA satellites and isolated field galaxies drives the model to maximally rely on halo assembly to explain environmental quenching (i.e., $r_c\sim 1$, Fig.~\ref{fig:rc_vmp} \& \ref{fig:rc}).
    
    \item UM-SAGA predicts that isolated field galaxies at $\mstar\sim 10^{6.5}\msun$ must quench at $f_Q\gtrsim 30\%$, under the model assumption that environmental quenching is purely attributed to halo assembly (Figs.~\ref{fig:fqm} and \ref{fig:rc}).
\end{enumerate}

Thus, UM-SAGA is an empirical galaxy--halo connection model that is self-consistently constrained by the latest low-mass galaxy observations down to $\mstar\gtrsim 10^7\msun$. The model provides a timely example of a self-consistent galaxy evolution framework, based on dark matter halo assembly histories, that is matched to the properties of 
satellites in a statistical sample of
Milky Way-mass host galaxies.  Despite its successes summarized above, our results indicate that exploration beyond the UM-SAGA framework is necessary to better understand low-mass galaxy formation. As discussed in Section~\ref{sec:discussion:lmc}, better characterizing halo assembly using halo properties that are sensitive to changes on different length and time scales will test whether the over-quenching of LMC-mass satellites predicted by UM-SAGA is caused by the specific choice of $\Delta v_{\rm max}$ in UM. Another issue discussed in Section~\ref{sec:results:distance-trend} is that the radial distribution of UM-SAGA quenched satellites is not as concentrated as in SAGA. Applying UM-SAGA to (sub)halo samples from high-resolution zoom-in simulations with more realistic subhalo accretion rates and disk disruption will help clarify this issue. 

Apart from the model augmentations regarding halo accretion and stripping, UM also lacks a model to account for reionization quenching~(e.g., \citealt{2000ApJ...539..517B,2001PhR...349..125B,2014ApJ...796...91B,2015ApJ...804..136W,2018ApJ...857...45T}), which is expected to impact ultra-faint dwarfs ($\mstar\lesssim10^{5}\msun$). Since UM currently only models star formation/quenching based on halo growth/stripping, a separate reionization model must be designed to capture reionization quenching at a specific redshift and halo mass scale. Although most UFDs in the MW~\citep{2014ApJ...796...91B,2014ApJ...789..147W,2014ApJ...789..148W,2015ApJ...804..136W,2021ApJ...920L..19S} seem to be quenched at $z>4$ ubiquitously by reionization (prior to their infall), recent observations of M31 UFDs show evidence of rejuvenated star-formation in Andromeda XIII~\citep{2023arXiv230513360S}. Therefore, such a reionization model should also be flexible enough to allow for an interplay between reionization quenching and the potential rejuvenation of SFR coupled to halo growth. The key focus of an upcoming follow-up paper will therefore be to jointly model LG low-mass galaxy SFHs, which probe the full evolution history of low-mass galaxies around the MW and M31, together with the larger sample of both Local Volume low-mass galaxies and the SAGA satellites and SDSS isolated field galaxies used in this work.

These efforts will help place the LG dwarfs in a broader cosmological context and shed light on the quantitative differences in satellite quenching between SAGA and LG environments. As a next step, we plan to re-calibrate  this reionization-UM model on the set of embedded disk simulations mentioned above and constrain the transition mass scale between environmental and reionization dominant quenching schemes. This will eventually lead to a unified framework for modeling galaxy star formation histories over the full range of halo masses that  host observable galaxies.

\medskip
The best-fit UM-SAGA model and source code repository are publicly available at \url{https://bitbucket.org/RW-Stanford/universemachine-saga/src/main/}. 

We thank Ralf Kaehler for generously sharing the multi-transparent-layer phase-space tessellation rendering code~\citep{KAEHLER201768,KAEHLER2018} to generate the projected dark matter density maps in Fig.~\ref{fig:model}. The HST data presented in this article were obtained from the Mikulski Archive for Space Telescopes (MAST) at the Space Telescope Science Institute. The specific observations analyzed can be accessed via MAST DOI \dataset[10.17909/jmnk-ky62]{https://doi.org/10.17909/jmnk-ky62}.

We thank Jenny Greene, Jiaxuan Li, Philip Mansfield, Daniel Weisz, and Andrew Wetzel for helpful discussions and insightful comments during the preparation of this draft. This research was supported by Heising-Simons Foundation grant 2019-1402, the National Science Foundation through NSF AST-1517148 and AST-1517422, NASA through HST-AR-17044, the U.S. Department of Energy under contract number DE-AC02-76SF00515 to SLAC National Accelerator Laboratory, and the Kavli Institute for Particle Astrophysics and Cosmology at Stanford University and SLAC. This research took place in part at the Kavli Institute for Theoretical Physics (KITP), supported by grant NSF PHY-1748958, and used computational resources at SLAC National Accelerator Laboratory, a U.S. Department of Energy Office and the Sherlock
cluster at the Stanford Research Computing Center (SRCC); the authors are thankful for the support of the SLAC and SRCC computational teams.

Support for Y.-Y.M.\ during 2019--2022 was in part provided by NASA through the NASA Hubble Fellowship grant no.\ HST-HF2-51441.001 awarded by the Space Telescope Science Institute, which is operated by the Association of Universities for Research in Astronomy, Incorporated, under NASA contract NAS5-26555.

This research used data from the SAGA Survey (Satellites Around Galactic Analogs; sagasurvey.org). The SAGA Survey is a galaxy redshift survey with spectroscopic data obtained by the SAGA Survey team with the Anglo-Australian Telescope, MMT Observatory, Palomar Observatory, W. M. Keck Observatory, and the South African Astronomical Observatory (SAAO). The SAGA Survey also made use of many public data sets, including: imaging data from the Sloan Digital Sky Survey (SDSS), the Dark Energy Survey (DES), the GALEX Survey, and the Dark Energy Spectroscopic Instrument (DESI) Legacy Imaging Surveys, which includes the Dark Energy Camera Legacy Survey (DECaLS), the Beijing-Arizona Sky Survey (BASS), and the Mayall z-band Legacy Survey (MzLS); redshift catalogs from SDSS, DESI, the Galaxy And Mass Assembly (GAMA) Survey, the Prism Multi-object Survey (PRIMUS), the VIMOS Public Extragalactic Redshift Survey (VIPERS), the WiggleZ Dark Energy Survey (WiggleZ), the 2dF Galaxy Redshift Survey (2dFGRS), the HectoMAP Redshift Survey, the HETDEX Source Catalog, the 6dF Galaxy Survey (6dFGS), the Hectospec Cluster Survey (HeCS), the Australian Dark Energy Survey (OzDES), the 2-degree Field Lensing Survey (2dFLenS), and the Las Campanas Redshift Survey (LCRS); HI data from the Arecibo Legacy Fast ALFA Survey (ALFALFA), the FAST all sky HI Survey (FASHI), and HI Parkes All-Sky Survey (HIPASS); and compiled data from the NASA-Sloan Atlas (NSA), the Siena Galaxy Atlas (SGA), the HyperLeda database, and the Extragalactic Distance Database (EDD). The SAGA Survey was supported in part by NSF collaborative grants AST-1517148 and AST-1517422 and Heising–Simons Foundation grant 2019-1402. SAGA Survey’s full acknowledgments are at \https{sagasurvey.org/ack}.

\software{
    Numpy \citep{numpy, 2020NumPy-Array},
    SciPy \citep{scipy, 2020SciPy-NMeth},
    Matplotlib \citep{matplotlib},
    IPython \citep{ipython},
    Jupyter \citep{jupyter},
    Astropy \citep{astropy}
}

\bibliographystyle{aasjournal}
\bibliography{main}

\appendix
\counterwithin{figure}{section}
\counterwithin{table}{section}

\section{UM-SAGA model posteriors}
\label{sec:App_A}

\begin{figure*}
    \centering
	\includegraphics[width=2\columnwidth]{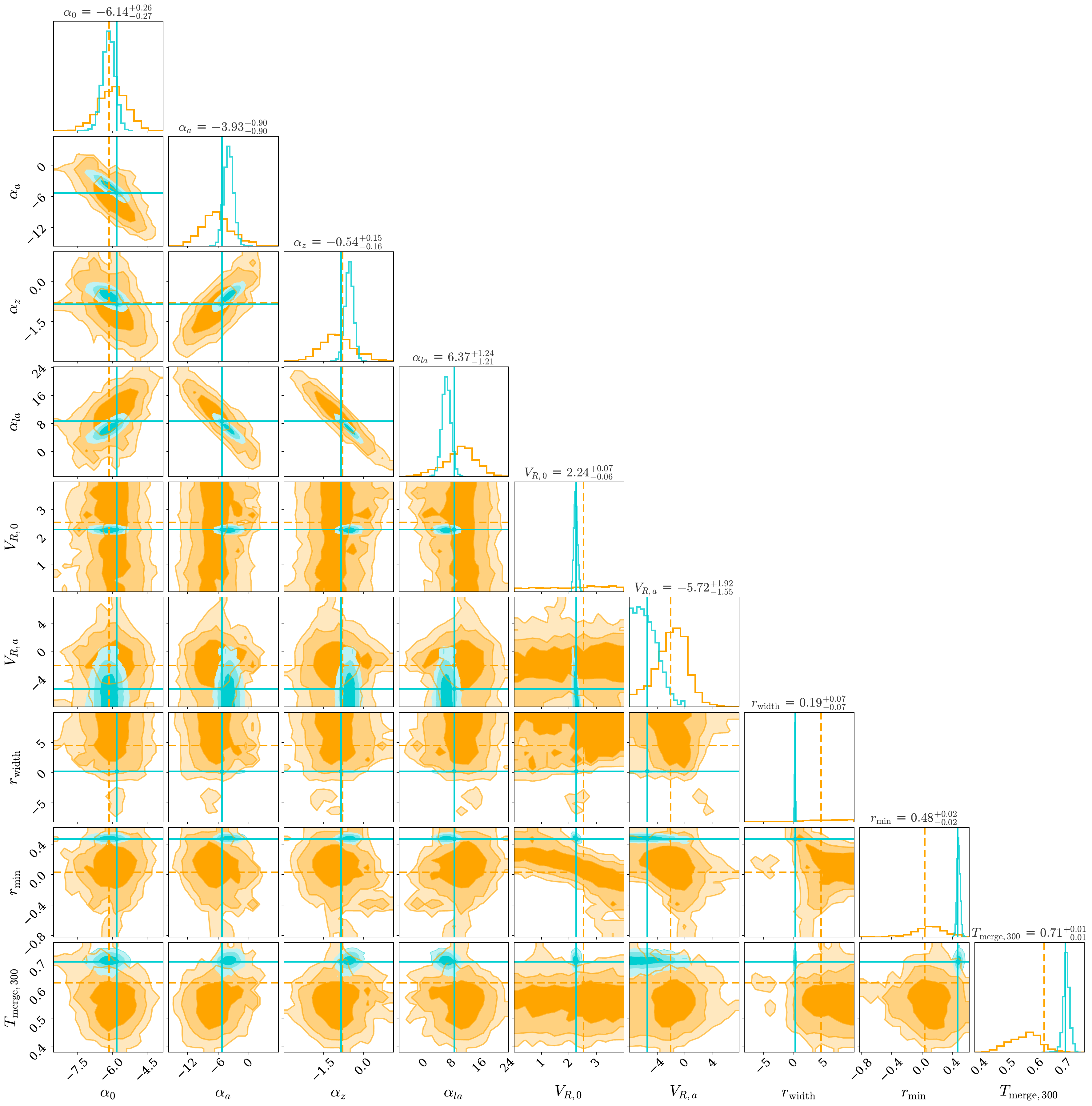}
    \caption{The posterior distribution for the nine parameters from UM DR1 related to low-mass galaxies. Orange shows the original posteriors of these parameters in UM DR1, and turquoise shows re-constrained posteriors in UM-SAGA. The diagonal subplots show the marginalized distributions for each parameter, with the labeled values denoting the median and $[16\%, 84\%]$ posterior. The off-diagonal subplots show the cross-correlations between each pair of parameters, with shaded contours marking out $68\%$, $95\%$, and $99\%$ distributions. The orange dashed line in each panel marks the UM DR1 best-fit value, while the turquoise solid line indicates the UM-SAGA best-fit value. The posteriors of parameters from the UM DR1 model are much more tightened in UM-SAGA with the addition of the low-mass galaxy observational constraints from SAGA and SDSS.}
    \label{fig:A1}
\end{figure*}

In Fig.~\ref{fig:A1}, we show the marginalized distributions of the nine parameters from UM DR1 explored in this work. Their priors, best-fit values, $95\%$ posteriors, and comparison to UM DR1 posteriors are presented in Table~\ref{tab:A}. The best-fit model is obtained by performing a least-$\chi^2$ fit of the model starting from the lowest-$\chi^2$ point in the MCMC samples, generally not equal to the median of each parameter's marginalized posterior median. All parameters are more tightly constrained in UM-SAGA than in UM DR1. 

The changes in the four low-mass end slope parameters in the mean $\mathrm{SFR}(v_{\rm Mpeak})$ relation are small compared to the scatter in their posteriors, but the low-mass end slope is much tighter constrained with the added low-mass galaxy data in this work. The changes in the four parameters related to $r_c$ ($V_{R,0}$, $V_{R,a}$, $r_{\rm min}$, $r_{\rm width}$) are the most drastic with the addition of the new low-mass galaxy constraints. From Table~\ref{tab:A}, $r_{\rm width}$ significantly decreases due to the sharp increase in $r_c$ required by the new quenched fractions from SAGA satellites and SDSS isolated field galaxies (Fig.~\ref{fig:rc_vmp}), while the $r_{\rm min}$ parameter also increases accordingly. The new low-mass galaxy constraints break the degeneracy between $V_{R,0}$ and $r_{\rm min}$ in UM DR1 due to only having auto-correlation constraints at one mass scale, $\mstar>10^{10.3}\msun$. Interestingly, $T_{\rm merge,300}$ increases from UM DR1 to UM-SAGA, requiring fewer orphans to be tracked; more galaxies are disrupted earlier to fit the SAGA data. We have checked that all parameters' MCMC chains converge to their posteriors (Gelman--Rubin statistics $\sim 1.2$). 

\begin{figure*}
    \centering
	\includegraphics[width=2\columnwidth]{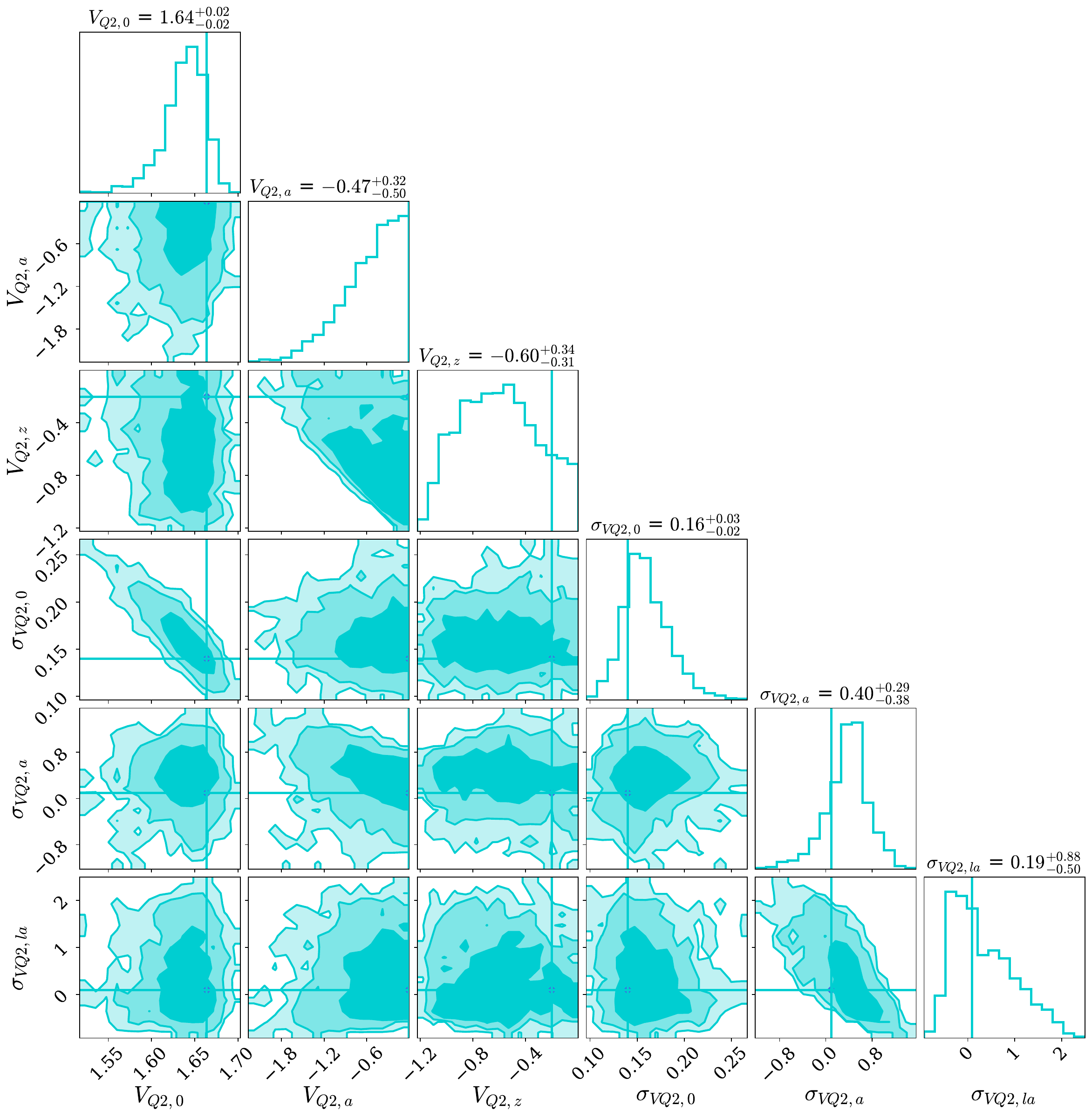}
    \caption{The posterior distribution for the six new low-mass quenching parameters added in this work. Legend is the same as Fig.~\ref{fig:A1}}
    \label{fig:A2}
\end{figure*}

\begin{figure}
    \centering
	\includegraphics[width=\columnwidth]{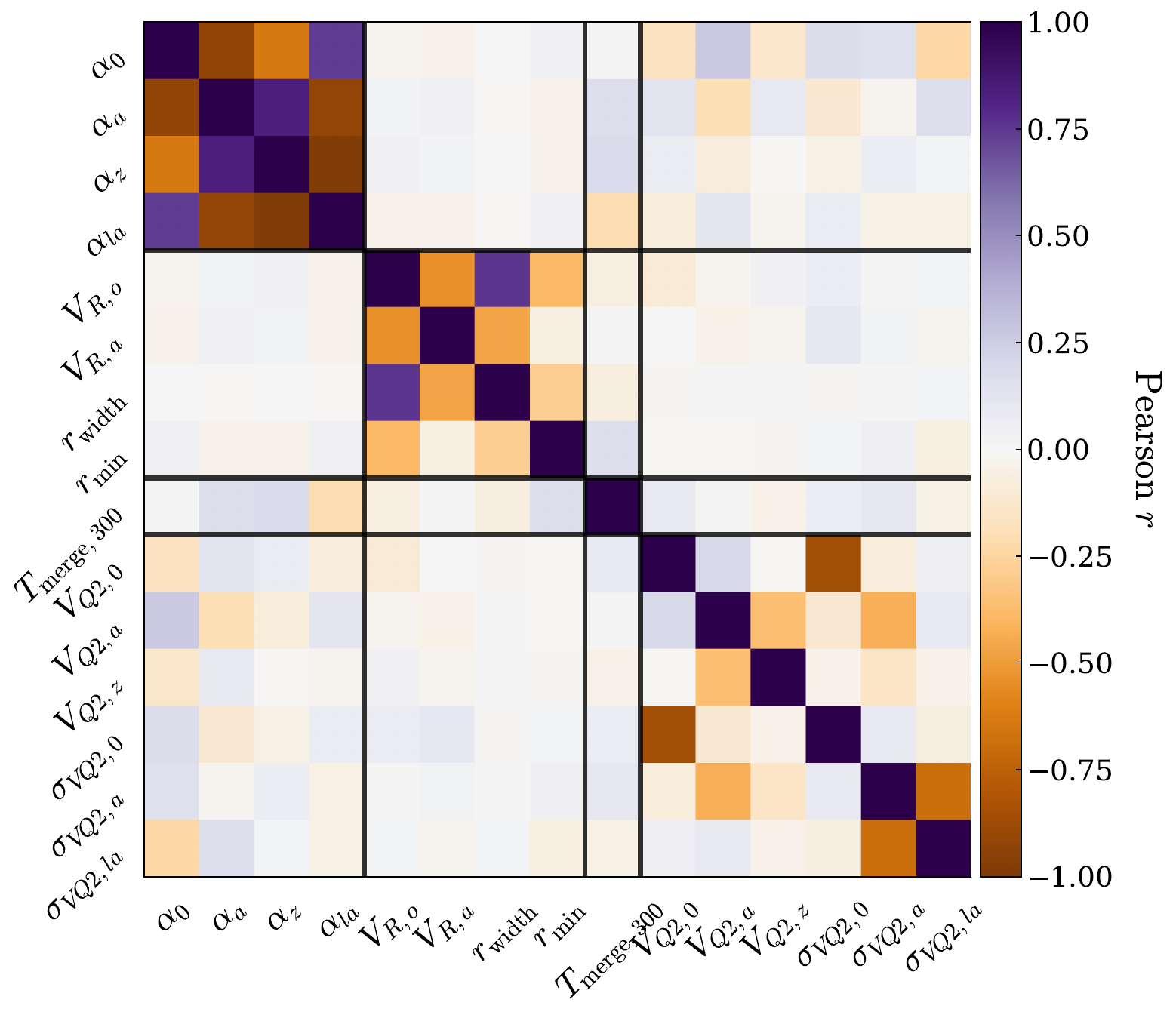}
    \caption{The model correlation matrix with the color bar showing the Pearson correlation coefficient. The thick lines mark major model blocks explored in this work: top left is the $\mathrm{SFR}-v_{\rm Mpeak}$ scaling relation; second to top left is the $r_{c}$ rank correlation between SFR and $\Delta (v_{\rm max})$ at fixed halo mass; the single parameter $T_{\rm merge, 300}$ is part of the orphan model; bottom right is the new low-mass quenching model added in this work. All major blocks have negligible cross-correlations, although some parameters share degeneracies within each block diagonal.}
    \label{fig:A3}
\end{figure}

In Fig.~\ref{fig:A2}, we show the marginalized distributions of the six new low-mass quenching parameters added in this work. $V_{Q2,0}$ and $\sigma_{VQ2,0}$ are the most constrained parameters by the new low-mass galaxy constraints at $z\sim 0$; they also show an apparent anti-correlation. This could be understood from the fact that most of the SAGA $f_Q$ points are at $\lesssim 0.5$ with the model being able to match the data equally well if the $f_Q=0.5$ mass scale ($V_{Q2,0}$) decreased while the scatter in the $f_Q(v_{\rm Mpeak})$ increased. It is also interesting to see that $\sigma_{VQ2, a}$ prefers a positive value even though no data is used to constrain the redshift evolution of the new low-mass quenching model, suggesting slower rises in $f_{Q}$ with decreasing (sub)halo mass towards higher redshifts just based on $z\sim0$ constraints. This could partially be due to the slight correlation between $V_{Q2,a}$ and the four $\alpha$ parameters as shown by the covariance matrix (Fig.~\ref{fig:A3}), but more careful tests in the future are required to further constrain the redshift evolution of the new low-mass quenching module using the SFHs of LG dwarfs.

If we apply the UM DR1 model to predict existing constraints and newly added low-mass galaxy observables, we obtain a model $\chi^2_{\rm DR1} = 972.72$. The new best-fit UM-SAGA model has a $\chi^2_{\rm new} = 402.71$ on the same data set. Since we have added $\Delta \nu = 6$ new parameters, and the number of degrees of freedom for the 1139 observational data points and 15-free parameter model is $\nu = 1124$, we calculate the $F$-statistics for the model update improvements:
\begin{equation}
    \label{eq:A1}
    F = \frac{\chi^2_{\rm DR1} - \chi^2_{\rm new}}{(\chi^2_{\rm new}/\nu) * \Delta\nu} = 265.16.
\end{equation}
If we assume the UM DR1 model had nine free parameters relevant for low-mass galaxies before adding the six low-mass quenching parameters in this work, then we obtain a $p$-value of $p = 1.33\times 10^{-9}$ for the $F$-test. This justifies that the reduction in $\chi^2$ due to the addition of the six low-mass quenching parameters is unlikely to be caused by the additional randomness in the new degrees of freedom but is rather due to the significant improvement in model flexibility of UM that made it more capable of matching the existing and newly added data constraints simultaneously. The correlation matrix of the 15 parameters explored in this work is presented in Fig.~\ref{fig:A3}. Most significant cross-correlations are within each module (block diagonal), especially for coefficients that account for each module's redshift evolution. The six new parameters in the low-mass quenching model are all weakly correlated with the nine UM DR1 parameters we explore.

\begin{table}
		\begin{center}
		\begin{tabular}{lccc}
			\hline
            \hline
			Parameter & UM-SAGA BF & UM DR1 BF & Prior\\
			\hline
            $\alpha_{0}$ & -6.17 & -6.14 & $U(-15, -1)$ \\
            $\alpha_{a}$ & -3.98 & -5.18 & $U(-15, -1)$\\
            $\alpha_{z}$ & -0.57 & -0.79 & $U(-15, -1)$\\
            $\alpha_{la}$ & 6.60 & 8.60 & $U(-15, -1)$\\
            $r_{\min}$ & 0.46 & 0.03 & $U(-1, 1)$\\
            $r_{\rm width}$ & 0.24 & 4.52 & $U(-10, 10)$\\
            $V_{R, 0}$ & 2.32 & 2.53 & $U(0, 4)$\\
            $V_{R, a}$ & -6.11 & -2.05 & $U(-8, 8)$\\
            $T_{\rm merge, 300}$ & 0.70 & 0.63 & $U(0.2, 1)$\\
            \hline
			$V_{Q2, 0}$ & 1.65 & - & $U(0, 2.23)$ \\
			$V_{Q2, a}$ & -0.26 & - & $U(-3, 0)$ \\
			$V_{Q2, z}$ & -0.75 & - & $U(-3, 0)$ \\
            $\sigma_{VQ2, 0}$ & 0.16 & - & $U(0.01, 2)$ \\
			$\sigma_{VQ2, a}$ & 0.39 & - & $U(-3, 3)$ \\
			$\sigma_{VQ2, z}$ & 0.08 & - & $U(-3, 3)$ \\
			\hline
		\end{tabular}
        \end{center}
		\caption{Summary of the best-fit (BF) values, marginalized median, and $[2.3\%, 97.7\%]$ posterior distributions for the UM-SAGA and UM DR1 models. We note that the priors shown here for the nine UM DR1 parameters are their original priors in \citet{2019MNRAS.488.3143B}; their UM-SAGA priors in this work are set to their UM DR1 posteriors. The upper half of the table shows the nine parameters we explore from UM DR1 and the bottom half shows the six new parameters we added in this work for low-mass quenching.}
		\label{tab:A}
\end{table}

\section{Comparison with UM DR1}
\label{sec:App_B}

\begin{figure}
    \centering
	\includegraphics[width=\columnwidth]{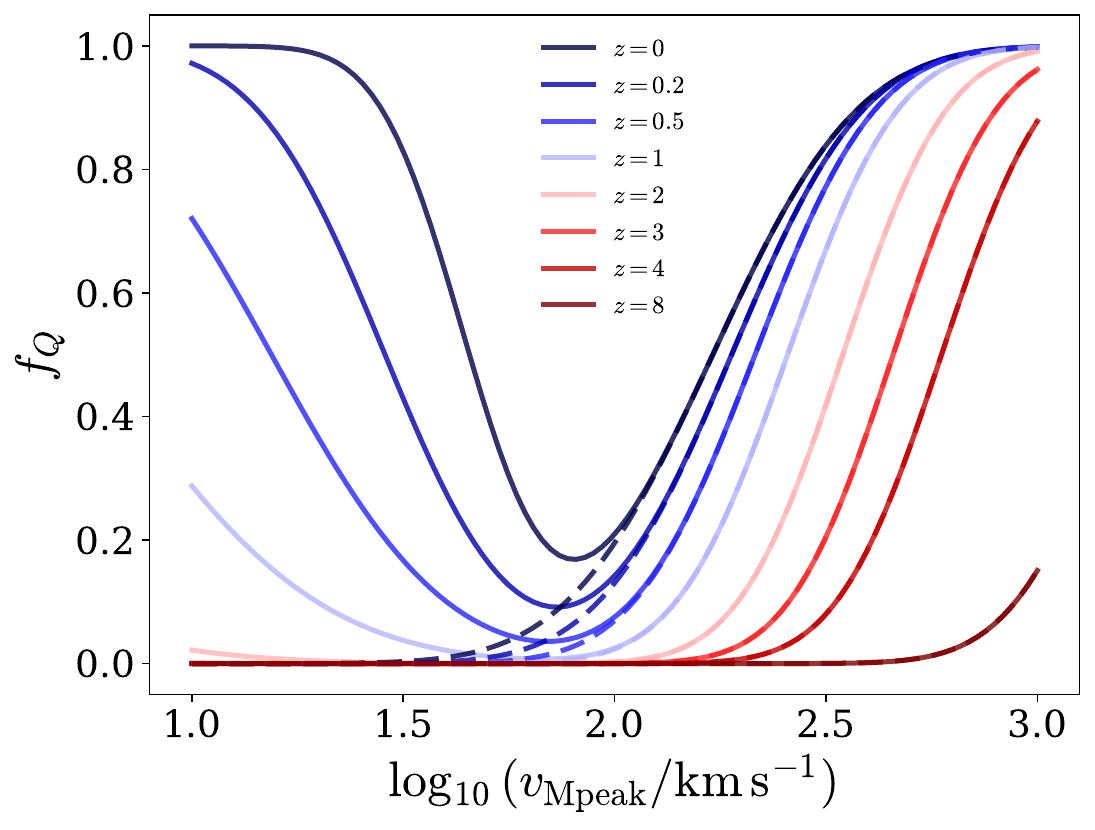}
    \caption{The $f_{Q}(v_{\rm Mpeak})$ functions of the best-fit UM-SAGA (solid) and UM DR1 (dashed) models. The different colors represent the model relations at different redshifts. The new low-mass quenching mostly affects halos with $v_{\rm Mpeak}\lesssim 100\mathrm{km\,s^{-1}}$.}
    \label{fig:B0}
\end{figure}

\begin{figure*}
    \centering
	\includegraphics[width=2\columnwidth]{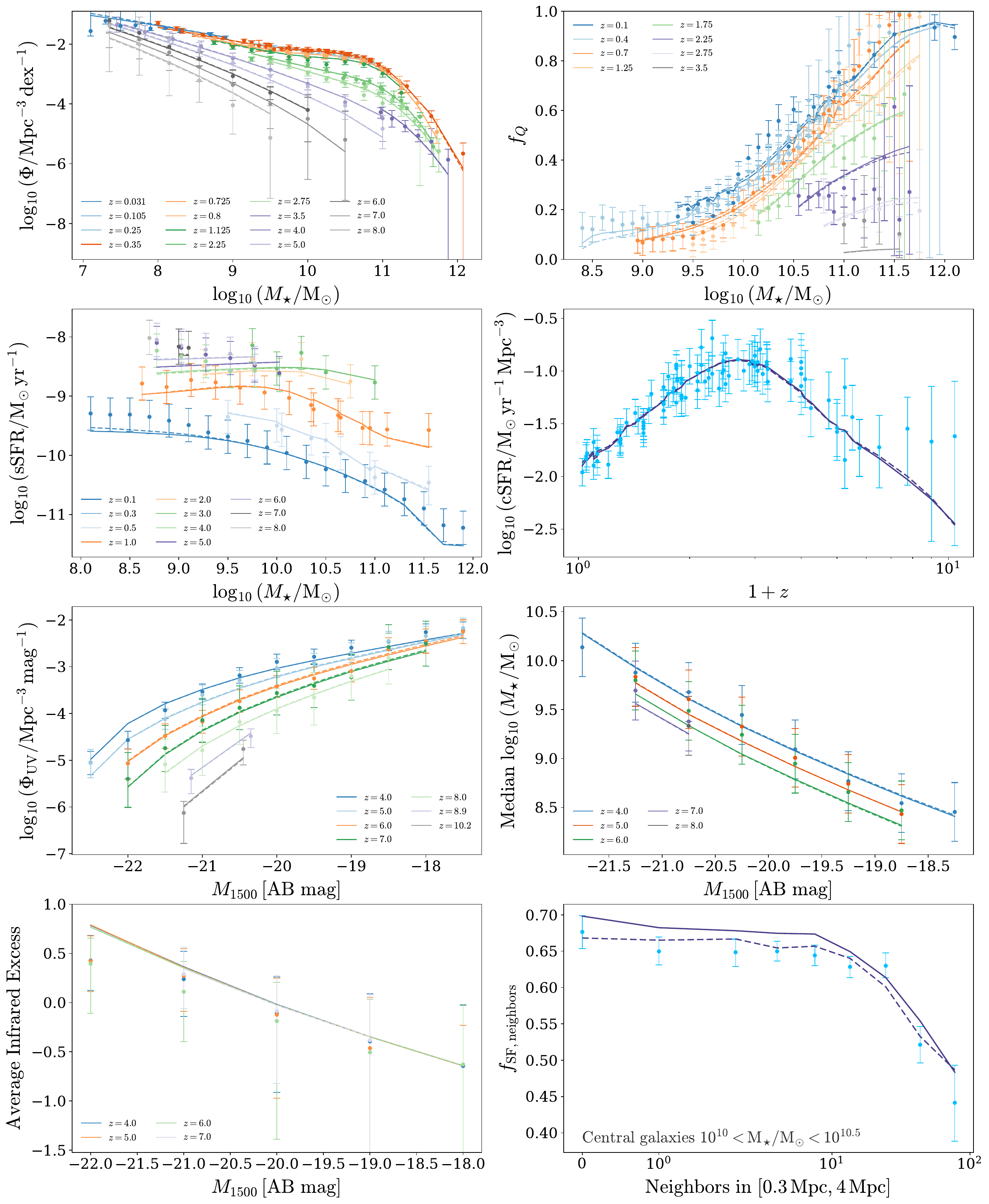}
    \caption{Part one of the model predictions by the UM-SAGA model (solid) compared to the UM DR1 model (dashed) for observables used in \citet{2019MNRAS.488.3143B}. {\it First row left panel}: stellar mass functions at various redshifts. {\it First row right panel}: overall quenched fraction of galaxies at various redshifts. {\it Second row left panel}: Mean specific SFR of galaxies at various redshifts. {\it Second row right panel}: Cosmic SFR density history. {\it Third row left panel}: UV luminosity function at $z\geqslant4$. $M_{1500}$ is the rest-frame absolute UV magnitude at 1500\AA. {\it Third row right panel}: Median stellar mass versus $M_{1500}$ at $z\geqslant 4$. {\it Fourth row left panel}: The average infrared excess (a probe of dust attenuation) at $z\geqslant 4$. {\it Fourth row right panel}: The star-forming fraction of neighboring galaxies as a function of neighbor counts from 0.3 Mpc to 4 Mpc around the central galaxy. Central galaxies with stellar masses $10^{10}<\mstar/\msun<10^{10.5}$ are considered. The UM-SAGA model predicts slightly higher $f_{\mathrm{SF, neighbors}}$ at neighbor counts $\lesssim 10$ than the UM DR1 model due to increased $r_c$. }
    \label{fig:B1}
\end{figure*}

\begin{figure*}
    \centering
	\includegraphics[width=2\columnwidth]{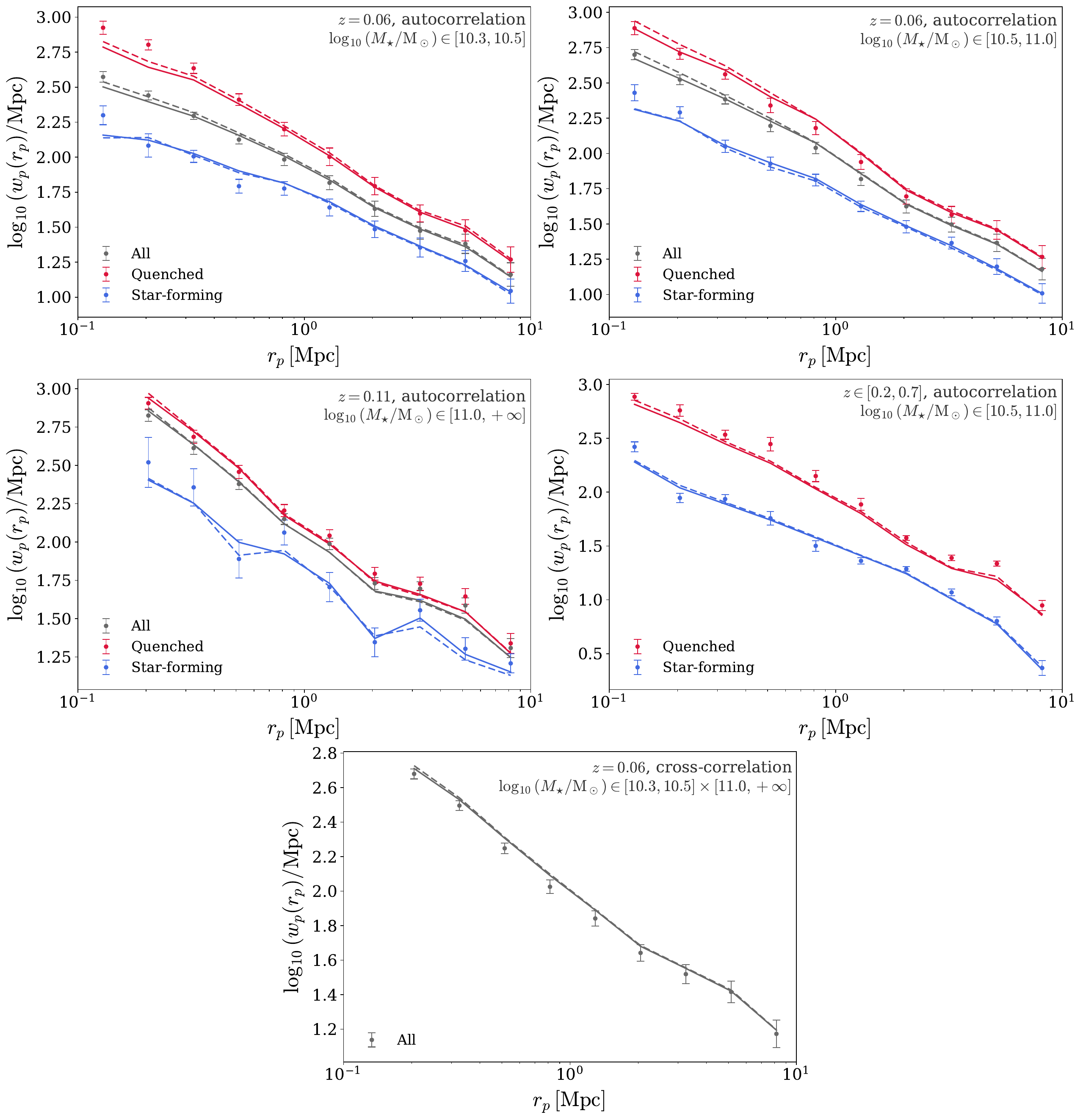}
    \caption{Part two of model predictions by the UM-SAGA model (solid) and the UM DR1 model (dashed) for galaxy clustering observables used in \citet{2019MNRAS.488.3143B}. The four panels in the first two rows show the projected auto-correlation functions for all galaxies (gray), quenched-only (red), and star-forming-only (blue) populations split into different stellar mass ($\mstar > 10^{10.3}\msun$) and redshift bins ($z<0.7$). The bottom panel shows the cross-correlation between a lower and higher stellar mass bin for all galaxies at a mean redshift of $z=0.06$. }
    \label{fig:B2}
\end{figure*}

In this appendix,  we show how the best-fit $f_{Q}(v_{\rm Mpeak})$ functions vary with redshift, comparing UM-SAGA to UM DR1 with the addition of the new low-mass quenching model in Fig.~\ref{fig:B0}. We also show the model predictions by the best-fit UM-SAGA model (this work) compared to the UM DR1 best-fit model~\citep{2019MNRAS.488.3143B} on all observational data constraints used in UM DR1 (Figs.~\ref{fig:B1} and \ref{fig:B2}). We show the \emph{shifted} observational values adapted for c125-2048 from \emph{Bolshoi-Planck} (see Section~\ref{sec:analysis:sim}). The takeaway message is that the UM-SAGA model still reproduces all observables used in UM DR1 well and that the joint re-calibration procedure for the nine UM DR1 parameters we explored does not significantly change the model predictions at higher mass scales ($\mstar \gtrsim 10^9 \msun$). The main difference in UM-SAGA is the slightly higher fraction of star-forming neighbors within [0.3, 4] Mpc from MW-mass hosts due to the slight increase in $r_c$ from UM DR1 to UM-SAGA at this mass range.

\section{Subhalo $v_{\rm Mpeak}$ and $\Delta v_{\rm max}$ distributions}
\label{sec:App_C}

\begin{figure*}
\begin{center}
    \includegraphics[width=2\columnwidth]{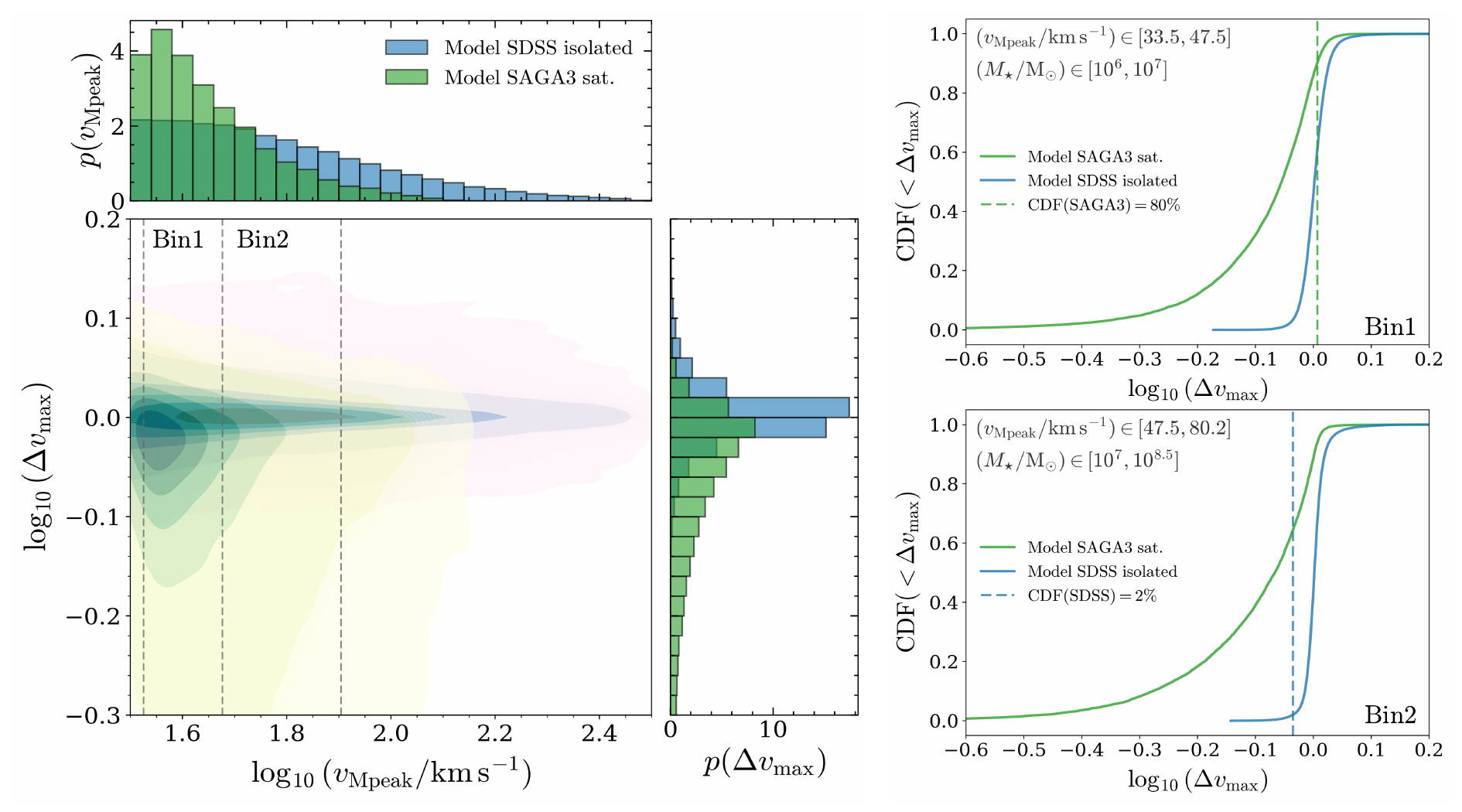}
    \caption{{\it Left panel}: The $v_{\rm Mpeak}$-$\Delta v_{\rm max}$ distribution of subhalos in c125-2048 that host SAGA-like satellites (green) or SDSS-like isolated field galaxies (blue). The side panels show the projected PDFs of $v_{\rm Mpeak}$ or $\Delta v_{\rm max}$ for the two low-mass (sub)halo populations. There is a significant overlap between the SAGA-like subhalos and isolated field halos in $\Delta v_{\rm max}$ at fixed $v_{\rm Mpeak}$. {\it Right panels}: CDFs of $\Delta v_{\rm max}$ in two $v_{\rm Mpeak}$ bins (marked by vertical dashed lines in the left panel) that roughly correspond to stellar mass ranges of $\mstar/\msun\in[10^6, 10^7]$ and $[10^7, 10^{8.5}]$. The CDFs for the isolated field halos in both bins are much steeper than the SAGA-like satellites occupying a narrower range of $\Delta v_{\rm max}$ in the same $v_{\rm Mpeak}$ bin. In the top right panel, we mark the $\Delta v_{\rm max}$ value at which the CDF of the SAGA-like subhalos reach $90\%$. This corresponds to simultaneously quenching at least $60\%$ of isolated field galaxies in $\mstar/\msun\in[10^6, 10^7]$ with $r_c\rightarrow 1$. In the bottom right panel, we mark the $\Delta v_{\rm max}$ value in the $\mstar/\msun\in[10^7, 10^{8.5}]$ bin where the CDF of the SDSS-like isolated field halos reach $2\%$ and that yields a maximum quenching fraction $~65\%$ for the SAGA satellites if $r_c = 1$. The actual SAGA quenched fraction (Fig.~\ref{fig:fqm}) is $\lesssim 60\%$ in this mass range. Thus the best-fit $r_c$ only needs to be $\sim 0.9$ (Fig.~\ref{fig:rc_vmp}). There are about $10\%$ of orphan galaxies in these two mass bins, which were not included in the CDFs.}
    \label{fig:C} 
\end{center}
\end{figure*}

This Appendix illustrates the limited difference in low-mass (sub)halo assembly histories for (sub)halos that host SAGA-like satellites and SDSS-like isolated field galaxies. This finite difference in (sub)halo properties is the fundamental cap preventing a larger $f_Q$ difference for SAGA satellites and isolated field galaxies when SFR is solely determined by halo accretion/stripping quantified via $\Delta v_{\rm max}$ in UM-SAGA.

To illustrate this, we show in the left panel of Fig.~\ref{fig:C} the $(v_{\rm Mpeak}, \Delta v_{\rm max})$ distribution of the \emph{(sub)halos} in the cosmological simulation c125-2048 that host low-mass galaxies either in the SAGA-like satellites sample or in the SDSS-like isolated field sample. As expected, the SAGA-like subhalos extend to much lower $\Delta v_{\rm max}<1$ due to more halo stripping in MW-mass host environments, while the majority of SDSS-like isolated field halos have $\Delta v_{\rm max}>1$ because they are steadily growing in mass in the field. The small amount of SAGA-like subhalos whose $\Delta v_{\rm max}>1$ are objects that have recently fallen into their hosts ($z_{\rm dyn} > z_{\rm Mpeak}$) and have not yet experienced significant tidal stripping, with their current $v_{\rm max}$ still higher than that from one dynamical timescale ago ($\sim 2$ Gyr). Conversely, a small portion of isolated field halos having $\Delta v_{\rm max}<1$ have recently been through major mergers, and their $v_{\rm max}$ has been declining in the past dynamical timescale due to (violent) relaxation. These two effects combined create a significant overlap in $\Delta v_{\rm max}$ for the two low-mass galaxy samples in different cosmic environments.  

Physically, it is reasonable to assign a higher SFR to low-mass satellites that have just been accreted onto their hosts without sufficient time to quench  (see quenching timescale constraints of dwarfs from \citealt{2013MNRAS.432..336W,2014MNRAS.442.1396W,2019MNRAS.483.4031R}), while also assigning a lower SFR to isolated field galaxies that have potentially quenched due to a recent major merger. Given the large overlap in $\Delta v_{\rm max}$ for bright dwarfs in different environments and the strong correlation ($r_c\rightarrow 1$) between halo stripping and galaxy SFR at $\mstar\lesssim 10^9\msun$, the $\delta f_Q \sim 40\%$ finite difference seen in the top panel of Fig.~\ref{fig:fqm} and left column of Fig.~\ref{fig:rc} represents an upper limit on the $f_Q$ difference between isolated field and MW-mass host environments if we assume low-mass galaxy quenching is only driven by differences in halo assembly. 

Indeed, as we show in the right panel of Fig.~\ref{fig:C}, one can find a $\Delta v_{\rm max}$ threshold in the $\mstar/\msun\in [10^7, 10^{8.5}]$ bin below which only $2\%$ of isolated field halos are quenched and still allowing for $\lesssim 65\%$ of SAGA-like satellites\ subhalos to quench, consistent with the large gap in $f_Q$ in the SAGA and SDSS data. This is why the best-fit UM-SAGA model works in this mass range, with a difference in satellite and isolated field quenched fraction of $~50\%$ achieved with $r_c < 1$. However, if we extrapolate UM-SAGA down to the $\mstar/\msun\in [10^6, 10^7]$ bin, and if we assume that $\sim 90\%$ of satellites in MW-mass hosts need to quench as in SAGA or the Local Group, then quenching based on $\Delta v_{\rm max}$ alone requires at $f_Q \sim 60\%$ in the isolated field. This is the reason why the maximum difference in satellite and isolated $f_Q$ shown in Fig.~\ref{fig:rc} is finite even in the case of $r_c\longrightarrow1$ and represents the lower limit required for field $f_Q$ if environmental quenching is purely determined by halo assembly.

\section{Example star formation histories for two low-mass galaxies}
\label{sec:App_D}

In this Appendix, we show the SFHs of four example galaxies in the stellar mass range $\mstar \in [10^{7.5}\msun, 10^{7.52}\msun]$ at $z=0$. Specifically, we show two SAGA-like satellites and two SDSS-like isolated galaxies. The SAGA-like satellites have lower $z=0$ SFRs and have built their stellar masses much earlier than their SDSS-like isolated counterparts, which have more prolonged and recent star formation.

\begin{figure}
    \centering
	\includegraphics[width=\columnwidth]{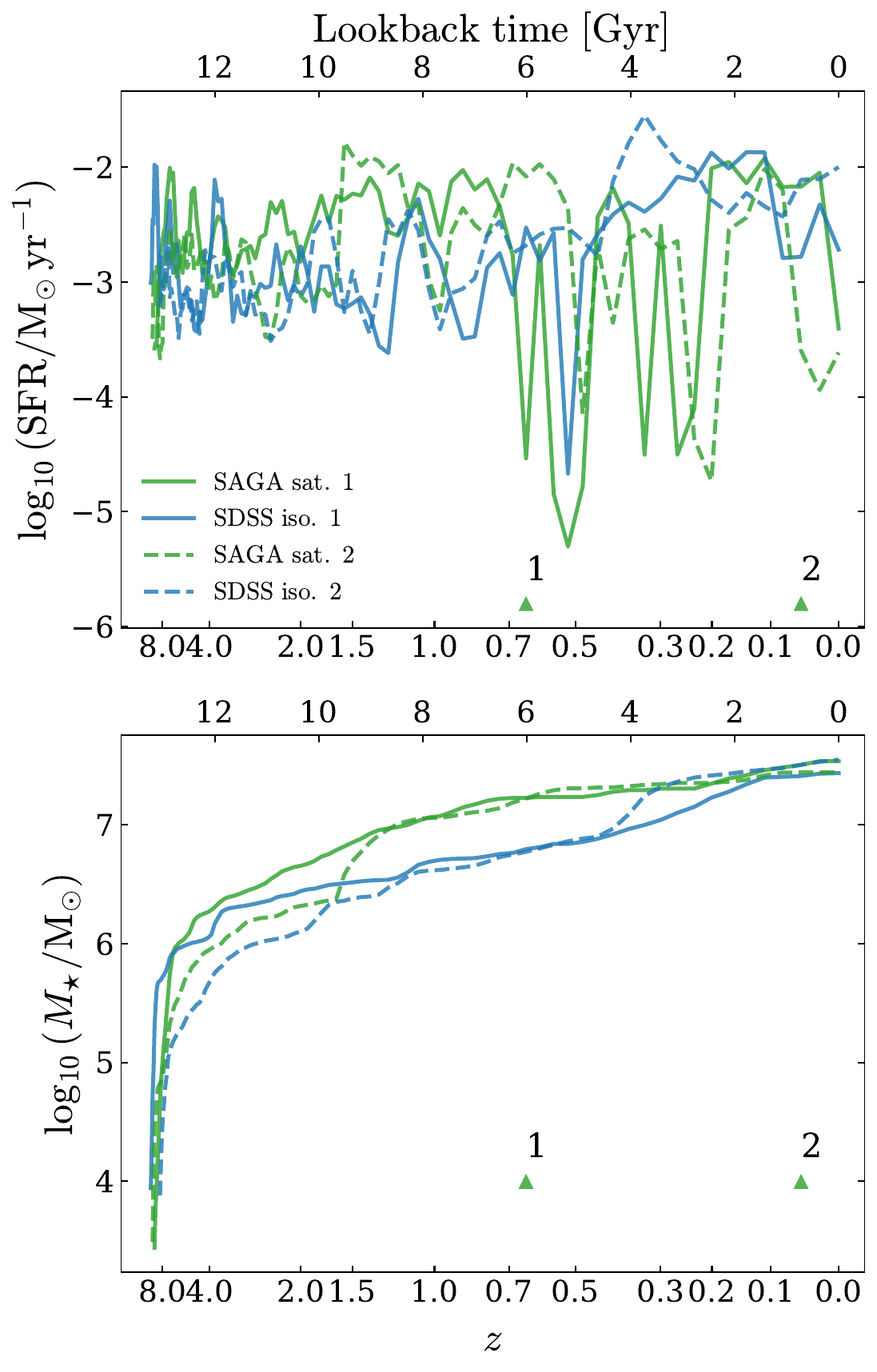}
    \caption{The star formation histories of four example low-mass galaxies at $\mstar \sim 10^{7.5}\msun$. The green are two SAGA-like satellites, and the blue are two SDSS-like isolated galaxies. The top panel shows their SFHs, and the bottom panel shows their stellar mass evolution. The green arrows indicate the two SAGA-like satellites' most recent accretion time. The stellar mass history takes into account the mass loss from the death of massive stars over time \citep[Eq. 21,][]{2019MNRAS.488.3143B}.}
    \label{fig:B-1}
\end{figure}

\end{document}